\documentclass[pre,preprintnumbers,amsmath,amssymb]{revtex4-1}
\usepackage{graphicx,color}
\usepackage{bm}
\usepackage{epsfig}
\usepackage{graphicx}
\usepackage{dcolumn}
\usepackage{bm}

\usepackage{lipsum}

\usepackage[FIGTOPCAP,tight,raggedright,nooneline]{subfigure}
\usepackage{float}
\graphicspath{{images/}{plots/}}

\renewcommand{\vec}[1]{\bm{#1}}

\begin{document}
\title{Position distribution in a generalised run and tumble process}
\author{David S. \surname{Dean}}
\affiliation{Univ. Bordeaux and CNRS, Laboratoire Ondes et
Mati\`ere  d'Aquitaine
(LOMA), UMR 5798, F-33400 Talence, France}
\author{Satya N. \surname{Majumdar}}
\affiliation{LPTMS, CNRS, Univ. Paris-Sud,
Universit\'e Paris-Saclay, 91405 Orsay, France}
\author{Hendrik \surname{Schawe}}
\affiliation{LPTM, UMR 8089, CY Cergy Paris Universit\'e,
CNRS, 95000 Cergy, France}

\bibliographystyle{apsrev}
\begin{abstract}
We study a class of stochastic processes of the type
$\frac{d^n x}{dt^n}= v_0\, \sigma(t)$ where $n>0$ is a positive integer and
$\sigma(t)=\pm 1$ represents an `active' telegraphic noise that flips from one state
to the other with a constant rate $\gamma$. For $n=1$, it
reduces to the standard run and tumble process
for active particles in one dimension. This process can be analytically continued
to any $n>0$ including non-integer values. We compute exactly the mean squared
displacement at time $t$ for all $n>0$ and show that at late times
while it grows as $\sim t^{2n-1}$
for $n>1/2$, it approaches a constant for $n<1/2$.
In the marginal case $n=1/2$, it grows very slowly with
time as $\sim \ln t$. Thus the process undergoes a {\em localisation}
transition at $n=1/2$. We also show that the position distribution $p_n(x,t)$
remains time-dependent even at late times for $n\ge 1/2$, but
approaches a stationary time-independent form for $n<1/2$.
The tails
of the position distribution at late times exhibit a large deviation form,
$p_n(x,t)\sim \exp\left[-\gamma\, t\,
\Phi_n\left(\frac{x}{x^*(t)}\right)\right]$, where $x^*(t)= v_0\, t^n/\Gamma(n+1)$.
We compute the rate function $\Phi_n(z)$ analytically for all $n>0$ and
also numerically using
importance sampling methods, finding excellent agreement between them.
For three special values $n=1$, $n=2$ and $n=1/2$ we compute the
exact cumulant generating function of the position distribution at all times $t$.
\end{abstract}
\maketitle

\date{\today}
\section{Introduction}

The position of an overdamped Brownian particle in one dimension
evolves with time via the stochastic Langevin equation
\begin{equation}
\frac{dx}{dt}= \sqrt{2D}\, \eta(t)\, ,
\label{Brown.1}
\end{equation}
where the friction coefficient is set to unity, $D$ represents the
diffusion constant and $\eta(t)$ is a zero mean Gaussian white noise
with correlator $\langle \eta(t)\eta(t')\rangle= \delta(t-t')$.
A natural generalisation of this process is a family of stochastic
processes indexed by a positive integer $n$~\cite{MSBC_1996}
\begin{equation}
\frac{d^nx}{dt^n}= \sqrt{2D}\, \eta(t)\, ,
\label{gen_lange.1}
\end{equation}
that reduces to the Brownian motion for $n=1$. For any $n> 1$,
the process $x(t)$, though simply Gaussian, is non-Markovian~\cite{MSBC_1996,BMS_2013}
due to the higher order derivative in \eqref{gen_lange.1}.

For $n=2$, \eqref{gen_lange.1}
represents the celebrated random acceleration process introduced by
Wang and Uhlenbeck~\cite{Wang_1945}, where an undamped
particle is subjected to a random force modelled
by the white noise. This problem
arises quite naturally in dispersion theory.
Consider a particle in a plane where its $y$-component undergoes
a Brownian motion $dy/dt= \eta(t)$, while in the
$x$-direction it gets convected by a noiseless flow velocity field $u(y(t))$
that depends only on the $y$-coordinate at time $t$, $dx/dt= u(y(t))$.
Thus, for a shear flow such that $u(y)= \gamma\, y$ (with $\gamma$
representing the shear rate), the $x$-component
undergoes the random acceleration process: $d^2x/dt^2= \gamma\, \eta(t)$.
In the mathematics literature, the process $x(t)$ in
\eqref{gen_lange.1} with $n=2$ has also been studied
extensively~\cite{Lachal_1997},
as it represents
the area under a Brownian curve
$x(t)= \int_0^t B(\tau) d\tau$ (where $B(t)$ represents a Brownian motion).
The random acceleration problem also has applications in
the context of granular collapse~\cite{collapse}.

For general $n$, an interesting application of \eqref{gen_lange.1}
can be found in the study of height fluctuations in equilibrium interface
models~\cite{Maj_Bray_01}. Here one considers the height profile $h(\vec x,t)$ of
an interface on a $d$-dimensional substrate of finite size, with
$\vec x$ denoting a point on the substrate.
At long times, the system reaches a stationary state where
the height profile, as a function of the spatial distance
along a fixed direction on the substrate, can be effectively
described by the process \eqref{gen_lange.1} (with $t$ denoting the
spatial distance and $x$ representing the height of the interface),
where the exponent $n$ can be expressed in terms of the dynamical
exponent of the interface~\cite{Maj_Bray_01,BMS_2013}.
Path integrals related to \eqref{gen_lange.1} with $n>1$
also arise in the treatment of semi-flexible polymers incorporating bending as well
as elastic energy \cite{klei86,bur93,dea19,pap77,smi01,kac13}, in higher
derivative field theories describing diblock copolymer phase separation
\cite{uch01,dea20}, and also in relativistic quantum mechanics \cite{sim90}.

The process \eqref{gen_lange.1} can be represented
(assuming for simplicity that all $(n-1)$ derivatives vanish at $t=0$)~\cite{MSBC_1996}
\begin{equation}
x(t)= \frac{\sqrt{2D}}{\Gamma(n)}\, \int_0^{t} ds\, (t-s)^{n-1}\, \eta(s)\,
\label{lin_rep.1}
\end{equation}
which reduces to \eqref{gen_lange.1} by repeated differentiation.
Using this representation, one can then analytically continue the
process even to non-integer (fractional) $n>0$.
Note that since $x(t)$ is a linear combination of Gaussian white noises, it is
clear that $x(t)$ is also a Gaussian process at all times $t$, with zero mean
and a variance that can be trivially computed from
\eqref{lin_rep.1} using the delta correlation of the noise
$\langle \eta(t)\eta(t')\rangle= \delta(t-t')$. One gets
\begin{equation}
\sigma_n^2(t)= \langle x^2(t)\rangle= \frac{2\, D\, t^{2n-1}}{(2n-1)\,
\Gamma^2(n)}\, .
\label{var_genw_n.1}
\end{equation}
Thus the position distribution, at any fixed time $t$, is given by
a purely Gaussian distribution
\begin{equation}
p_n(x,t)= \sqrt{\frac{(2n-1)\,
\Gamma^2(n)}{4\, \pi\, D\, t^{2n-1}}}\, \exp\left[-
\frac{(2n-1)\,\Gamma^2(n)}{4\,D\, t^{2n-1}}\, x^2\right]\,
\label{pnxt_intro.1}
\end{equation}
valid for all $x$, all $t$ and all $n>1/2$. Note that for $0<n<1/2$,
the process \eqref{lin_rep.1} is not well defined, because at short times,
the particle gets an infinitely large kick by the noise which sends it
to $\pm \infty$ leading to a pathological situation.

Finally, let us remark that even though the position distribution for
the process \eqref{lin_rep.1} is trivially Gaussian at all times for all
$n>1/2$, the first-passage properties of the process for any $n\ne 1$
is highly nontrivial due to non-Markovian nature
of the process~\cite{MSBC_1996,BMS_2013}. Even for $n=2$, it took almost
$47$ years, since
the original introduction of the problem by Wang and Uhlenbeck in 1945, to
compute the first-passage probability~\cite{bur93,sin92,Burkhardt_2016}.
The first-passage properties for general
$n>1/2$ have been studied extensively in recent
times by various methods and are again nontrivial
due to non-Markovian nature of the process for
$n\ne 1$~\cite{MSBC_1996,Maj_1999,SM_2007,BMS_2013,PS_2018}.

Another non-Markovian generalisation of the ordinary Brownian motion in
\eqref{Brown.1} that has been studied extensively is the so called
`persistent Brownian motion' ~\cite{kac74,mas96,wei02,mas17}
\begin{equation}
\frac{dx}{dt}= v_0\, \sigma(t)\, ,
\label{pbm.1}
\end{equation}
where $v_0$ denotes the intrinsic speed of a particle and the
noise $\sigma(t)$ is telegraphic: it can take two values $\pm 1$.
It flips from
one value to the other with a finite rate $\gamma$. This
model has seen a recent resurgence of interest in the
context of `run and tumble' (RTP) dynamics of an active particle
like the E. Coli bacteria~\cite{ber14,tai08}.
In one dimension, when the noise $\sigma(t)$
remains unflipped for a certain duration, the particle `runs' with
speed $v_0$ in that direction. When $\sigma(t)$ changes sign, it represents
a `tumble' and the particle changes its direction of motion and goes for
another run and so on.

The effective driving noise $\xi(t)=v_0\, \sigma(t)$
in \eqref{pbm.1} is `coloured' since its autocorrelation function (see
Section IV for a simple derivation)
\begin{equation}
\langle \xi(t) \xi(t')\rangle= v_0^2\, e^{-2\, \gamma\, |t-t'|}\, .
\label{autocorr.1}
\end{equation}
has a finite persistence time $\sim \gamma^{-1}$. The noise thus
has a memory of finite duration which makes the process $x(t)$
non-Markovian. In the limit $\gamma\to \infty$,
$v_0\to \infty$ but keeping the ratio $D= v_0^2/{2\gamma}$ fixed,
the noise $\xi(t)$ reduces to
a white noise since
\begin{equation}
\langle \xi(t) \xi(t')\rangle=
\frac{v_0^2}{\gamma}\, \left[\gamma\, e^{-2\gamma|t-t'|}\right]
\to 2D\, \delta(t-t')\, .
\label{autocorr.2}
\end{equation}
Thus the RTP dynamics \eqref{pbm.1} reduces to an ordinary Brownian
motion in this diffusive limit.

Several properties of the one dimensional
RTP process \eqref{pbm.1} such as the
position
distribution~\cite{wei02,HV_2010,ODA_1988,MADB_2012,Malakar_2018,EM_2018,Dhar_2019,SBS_2020} and first-passage
properties~\cite{ang14,ang15,Malakar_2018,EM_2018,led19,sin19,led20,ban20,mor20} are well known.
For example, for a particle starting
from $x=0$ at $t=0$ with $\sigma(0)=\pm 1$ with equal probability,
the position distribution $p_1(x,t)$ at finite $t$ is highly
nontrivial. The distribution is supported over the
interval $x\in [-v_0\, t, v_0\, t]$ and is given for $|x|\le v_0\, t$ by
\begin{equation}
p_1(x,t) =\frac{{\rm e}^{-\gamma t}}{2} \left\{ \delta(x-v_0 t)
+\delta(x+v_0 t)
+ \frac{\gamma}{2v_0}\left[ I_0(\rho) +
\frac{\gamma I_1(\rho)}{\rho}\right] \theta(v_0t-|x|)
\right\}\, ; \quad {\rm where}\quad
\rho = \sqrt{v_0^2 t^2 - x^2}\,\frac{\gamma}{v_0}
\label{p1xt.1}
\end{equation}
and $I_0(\rho)$ and $I_1(\rho)$ are modified Bessel functions of the
first kind. The edges of the support $x=\pm v_0\, t$ corresponds to
the maximal possible displacements of the particle on either sides
of the origin (corresponding to the event when the noise $\sigma(t)$ does not flip sign at all during time $t$).
As time progresses, the centres of the
two delta functions at the two edges
move ballistically away with speed $\pm v_0$ (representing
two {\em light cones}), but their amplitudes
decay exponentially with time $t$ since the
probability that $\sigma(t)$ retains its sign up to $t$ decays
as $\sim e^{-\gamma\, t}$. In the central part near $x=0$,
the distribution $p_1(x,t)$ approaches a Gaussian form at late times,
as one would expect since the RTP at late times does reduce to
the ordinary Brownian motion. These features of $p_1(x,t)$ are
well captured by a large deviation form exhibited by the distribution.
To see this, consider
the limit $t\to \infty$, $|x|\to \infty$
but with the ratio $z=x/{v_0\, t}$ fixed. Using the asymptotic behavior
$I_m(\rho)\sim e^{\rho}/{\sqrt{2\,\pi\, \rho}}$ as $\rho\to \infty$ for any $m>0$,
one finds that $p_1(x,t)$
in \eqref{p1xt.1} exhibits the following large deviation form
\begin{equation}
p_1(x,t) \sim \exp\left[-\gamma\, t\, \Phi_1\left(\frac{x}{v_0 t}\right)\right]\, \quad {\rm with}\quad
\Phi_1(z)= 1- \sqrt{1-z^2} \, , \,\,\,\, -1\le z\le 1 \, .
\label{phi1z_intro.1}
\end{equation}
The rate function behaves quadratically $\Phi_1(z)\approx z^2/2$ as $z\to 0$.
Substituting this behaviour in \eqref{phi1z_intro.1}, one finds that for $|x|\ll v_0\, t$,
the distribution converges to the Gaussian form,
$p_1(x,t)\sim \exp[-x^2/{4Dt}]$, with $D= v_0^2/{2\gamma}$, as expected for a Brownian motion
with diffusion constant $D$. Thus the large deviation regime valid at the tails, when extrapolated
towards the peak, matches smoothly with the inner
Gaussian peak characterizing the typical fluctuations around the mean.

Given that a finite memory encoded in the driving telegraphic noise
induces a nontrivial position distribution $p_1(x,t)$ at finite $t$ for the RTP
process \eqref{pbm.1}, it is natural to investigate the similar question
for the generalised RTP process that we introduce here
\begin{equation}
\frac{d^n x}{dt^n}= v_0\, \sigma(t)\, ,
\label{dnx_tel.1}
\end{equation}
where $n>0$ is a positive integer and $\sigma(t)=\pm 1$ is the
dichotomous telegraphic noise. For $n=1$, this process reduces
to the standard RTP \eqref{pbm.1}. For general $n$, this is an
`active' counterpart
of the `passive' white noise driven process \eqref{gen_lange.1}. The activeness
arises from the
driving noise being `coloured', i.e., with a finite memory encoded
in the persistence time $\gamma^{-1}$. Once again, we can use the
integral representation
\begin{equation}
x(t)= \frac{v_0}{\Gamma(n)}\, \int_0^{t} ds\, (t-s)^{n-1}\, \sigma(s)\,
\label{lin_rep.2}
\end{equation}
to define the process for any $n>0$, including non-integer values.
In this paper, our main goal is to simply study the position distribution
$p_n(x,t)$ of the process \eqref{lin_rep.2} for general $n>0$.

\begin{figure}
\includegraphics[width=1\textwidth]{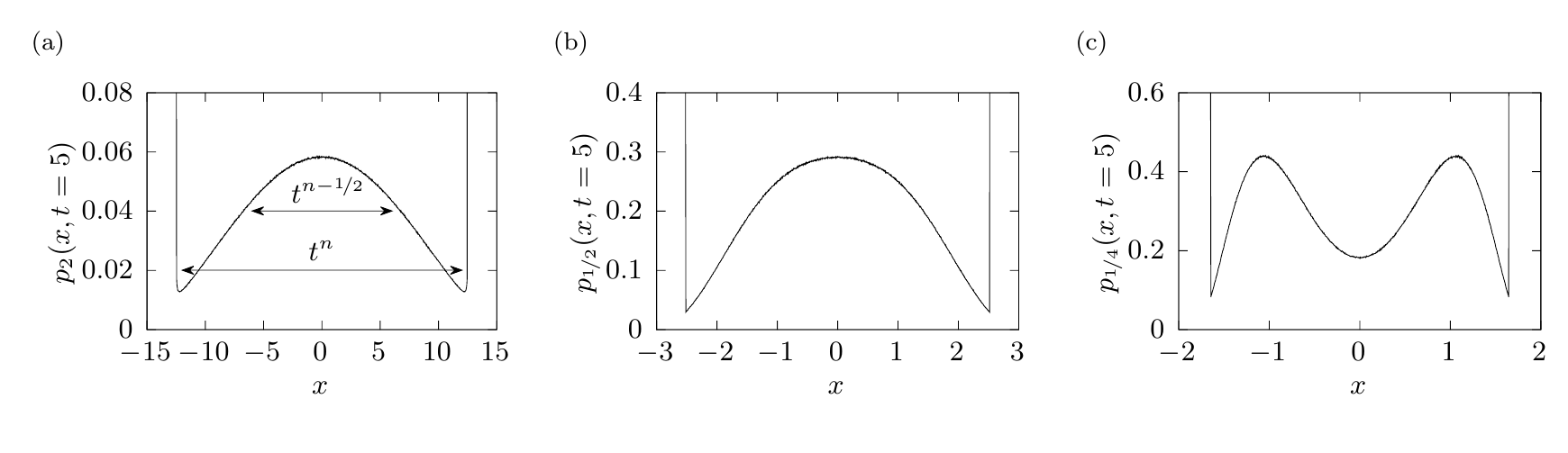}
  \caption{
    The position distribution $p_n(x,t)$ vs. $x$ for a fixed $t=5$ obtained from
    simulations for three different values of $n$: (a) $n=2$ (b) $n=1/2$ and
    (c) $n=1/4$.
    In all cases, the distribution has a support on $x\in [-x^*(t),x^*(t)]$ where
    $x^*(t) = v_0\, t^n/\Gamma(n+1)$.
    In (a), where $n>1/2$, the typical fluctuations grow
    as $\sim t^{n-1/2}$ at late times
    and the distribution on this scale approaches a Gaussian form (near the peak).
    In (c), where $n<1/2$, the typical fluctuations are of $O(1)$ at late times
    and the distribution approaches a time-independent double-humped
    structure at late times.
    The case (b) where $n=1/2$ is the marginal case, separating
    the `localised' phase ($n<1/2$) and the `de-localised' phase ($n>1/2$),
    where the typical
    fluctuations grow logarithmically at late times and
    the distribution remains time-dependent
    for large $t$. In all the three cases, the distribution
    on the large deviation scale,
    when $x\sim x^*(t)$, satisfies the large deviation
    form in \eqref{ldt_intro.1}.
}
\label{fig:schematic}
\end{figure}

Let us highlight briefly our main results. First, we show that
the distribution $p_n(x,t)$ of the process \eqref{lin_rep.2}
is well defined for any $n>0$, unlike the white noise driven process
\eqref{lin_rep.1} which is pathological for $0<n<1/2$.
Secondly, we find a very interesting {\em localisation transition}
at the critical value $n=1/2$. While the mean squared displacement
of the particle increases for large $t$ as $\sim t^{2n-1}$ for
$n>1/2$, it approaches a constant as $t\to \infty$ for $n<1/2$.
In the marginal case $n=1/2$, the mean squared displacement grows
logarithmically as $\sim \ln t$ at late times. This result is
proved analytically by computing the mean squared displacement exactly for
all $t$. This localisation transition also shows up in the
full distribution $p_n(x,t)$. We show that for $n< 1/2$, the
distribution $p_n(x,t)$ at late times approaches a time-independent form
for $|x|\ll t^n$. Moreover, this stationary distribution for $n<1/2$
has a {\em double-humped} structure (see Fig.~\ref{fig:schematic}).
In contrast, for $n\ge 1/2$, the distribution at late times
remains time-dependent even for $|x|\ll t^n$ and near
its peak at $x=0$ it approaches a Gaussian form (see Fig.~\ref{fig:schematic}), as
in \eqref{pnxt_intro.1} for the white noise driven process. The case
$n=1/2$ is a marginal one, where the typical fluctuations grow as $\ln t$ at late times and
the distribution remains time-dependent even at late times.

In this paper, we also demonstrate analytically that
the distribution $p_n(x,t)$, supported
inside the light cone $x\in [-x^*(t), x^*(t)]$
with $x^*(t)=v_0\, t^n/\Gamma(n+1)$, exhibits a large deviation
behaviour
\begin{equation}
p_n(x,t)\sim \exp\left[-\gamma\, t\, \Phi_n\left(\frac{x}{x^*(t)}\right)
\right]\, .
\label{ldt_intro.1}
\end{equation}
We compute the large deviation rate function $\Phi_n(z)$ analytically for all
$n>0$ which recovers \eqref{phi1z_intro.1} for $n=1$.
The large deviation form in \eqref{ldt_intro.1} holds when $x\sim x^*(t)\sim t^{n}$.
For $n\ge 1/2$, \eqref{ldt_intro.1}
turns out to hold for much smaller $|x|$, i.e., even when $|x|\sim t^{n-1/2}$ and indeed matches
smoothly with the central Gaussian peak describing typical fluctuations.
In contrast, for $n<1/2$, Eq. (\ref{ldt_intro.1})
only describes the behaviour
of the distribution near the two light cones $x\sim \pm x^*(t)$, but
does not describe the stationary double-humped structure near the centre. It does predict however
that for $0<n<1/2$, the stationary distribution at late times has super-exponential tails,
$-\ln p_n(x, t\to \infty)\sim |x|^{1/n}$ as $|x|\to \infty$.
For convenience, we provide a summary of our main results and formulae in
Section II.

Thus one of our main findings for the process \eqref{lin_rep.2}
is the emergence of a localised phase for $0<n<1/2$ due to
the finite memory of the driving noise.
This localised phase induced by the active telegraphic noise
has no analogue in the passive white noise driven process
in \eqref{lin_rep.1}. One may wonder if there is
a physical system that corresponds to this new localised
phase for $0<n<1/2$. Indeed there is an interesting physical
system that corresponds precisely to this $0<n<1/2$ case, as we briefly discuss
now.

We consider an elastic interface of height $h({\bf x},t)$
on a $d$-dimensional substrate. An external time-dependent
force $f(t)$ is applied locally, say at the origin.
The energy of a height configuration is then given by
\begin{equation}
E[\{h({\bf x},t)\}]=
\int d{\bf x}\,\left[\frac{\kappa}{2}\, (\nabla h)^2 -f(t)\,
h({\bf x})\, \delta({\bf x})\right]\, .
\label{energy.1}
\end{equation}
The first term represents the surface energy with stiffness
$\kappa>0$ and
the second term is due to the applied force at the origin.
Consider the generic zero temperature (noiseless) dynamics
of the interface~\cite{Maj_Bray_01,BMS_2013}
\begin{equation}
\frac{\partial h({\bf x},t)}{\partial t} =
-[ -\nabla^2]^{\frac{z_D}{2}}\frac{\delta E[h]}{\delta h}=
\left[ -\nabla^2\right]^{\frac{z_D}{2}}
\left[\kappa\, \nabla^2 h({\bf x},t)+f(t)\delta({\bf x})\right]\, ,
\label{inter_dyn.1}
\end{equation}
where $z_D\ge 0$ is an exponent that parametrizes the dynamics.
For example, $z_D=0$ for the standard model A type dynamics~\cite{bra95},
while $z_D=2$ for the model B dynamics \cite{bra95}
where the total height is conserved.
At finite temperature, one may add a thermal additive noise
on the right hand side of \eqref{inter_dyn.1}, but we
restrict here to zero temperature for simplicity.
Since \eqref{inter_dyn.1}
is linear in $h$, it can be solved using the spatial Fourier transform
of the height: ${\tilde h}({\bf k},t) =
\int_{-\infty}^\infty d{\bf x}\,
\exp(-i{\bf k}\cdot{\bf x})\,
h({\bf x},t)$. Taking the Fourier transform of \eqref{inter_dyn.1} gives
\begin{equation}
\frac{\partial \tilde h({\bf k},t)}{\partial t}
= -\kappa\, |k|^{2+z_D}\, \tilde h({\bf k},t) + f(t)\, |k|^{z_D}\, .
\label{hkt_dyn.1}
\end{equation}

Assuming we start from a flat initial condition $h({\bf x},0)=0$, one
finds the explicit solution for general force $f(t)$ as
\begin{equation}
\tilde h({\bf k},t)= \int_0^t
ds\, \exp\left(-\kappa\, (t-s)\, |k|^{2+z_D}\right)\, |k|^{z_D}\, f(s).
\label{hkt_sol.1}
\end{equation}
The position $h(0,t)$ of the point where the force is applied is then given by
\begin{eqnarray}
h(0,t) &=&\frac{1}{(2\pi)^d}\int d{\bf k}
\int_0^t ds \exp\left(-\kappa\, (t-s)\, |k|^{z_D+2}\right)\,|k|^{z_D}\,
f(s)\nonumber \\
&=&\frac{d\,\Gamma(\frac{d+z_D}{2+z_D})}{(2+z_D)\,
\Gamma(\frac{d}{2}+1)\,2^d\, \pi^{\frac{d}{2}}\,\kappa^{\frac{d+z_D}{2+z_D}}}\,
\int_0^t ds\  (t-s)^{-\frac{d+z_D}{2+z_D}}\, f(s) \, .
\label{h0t_sol.1}
\end{eqnarray}
Consider now the external force to be noisy telegraphic, i.e.,
$f(t)=v_0\, \sigma(t)$. Comparing with \eqref{lin_rep.2}, we see
the correspondence
\begin{equation}
n = \frac{2-d}{2+z_D} \, .
\label{n_corr.1}
\end{equation}
For any $z_D\ge 0$, clearly $n<1$ in \eqref{n_corr.1} and
we thus have a physical realization of the process \eqref{lin_rep.2}
with $n<1$. By tuning the exponent $z_D$ (which corresponds to
choosing different dynamics of the interface), one can physically realize
various values of $n<1$ of our process in \eqref{lin_rep.2}.

We also note from \eqref{n_corr.1} that since the process \eqref{lin_rep.2} is
well defined only for $n>0$, we must have $d<2$. Thus the most relevant `physical'
dimension is $d=1$, where $n=1/(2+z_D)$.
Indeed, in $d=1$, similar activity driven interface models
have been studied recently, in the context of a Rouse
polymer chain in the presence of both the Gaussian white noise and
the telegraphic noise~\cite{osm17,cha19}. These models, in the continuum limit of long chains,
correspond precisely to our model with $z_D=0$ (model-A dynamics), and hence $n=1/2$.
Similarly, by choosing $z_D=2$ in the interface model (model-B dynamics), we can realize
$n=1/4$. The choice $z_D=1$, which applies to the effective dynamics of an interface separating two
phases each of which is undergoing model-B dynamics, would correspond to $n=1/3$ in
our process.

The rest of our paper is organised as follows. In
Section \ref{summary} we provide a quick summary
of our main results with the relevant formulae. In Section \ref{trick},
using a simple trick we
provide a different representation of the process \eqref{lin_rep.2}
that allows us to compute the mean squared displacement as well
as the one point position distribution $p_n(x,t)$ in a
relatively easier fashion.
In Section IV, we compute the mean squared displacement
explicitly for all $n>0$ that already indicates the existence
of the localisation transition at $n=1/2$.
In Section V, we derive an exact Feynman-Kac evolution equation
for the cumulant generating function of the position distribution.
In Section VI, we show how to
compute the large deviation behaviour of the
cumulant generating function at late times. In Section VII,
we extract the large deviation behaviour of $p_n(x,t)$ from
that of the cumulant generating function.
In Section \ref{numerics} we confirm our analytical results via numerical
simulations, and in particular we discuss the importance sampling method that
we use to compute the rate functions characterising the large deviations.
Finally we provide some concluding remarks
in Section IX.
Details of the exact solution of the cumulant generating function for the
three special cases $n=1$, $n=2$ and $n=1/2$ are provided in the Appendix.

\section{Summary of Main Results}
\label{summary}

Since the paper is a bit long, it is perhaps useful, for the convenience of the readers,
to provide a brief summary of our main results along with the relevant formulae so that they
are easily retrievable if needed. This section does precisely that and the actual
derivations are provided in later sections and in the Appendix.

\vskip 0.3cm

{\noindent {\bf Mean squared displacement:}} For the process \eqref{lin_rep.2} with arbitrary $n>0$,
while the mean position vanishes identically, $\langle x(t)\rangle=0$ (given $x(0)=0$), the variance
or the mean squared displacement is a nontrivial function of $t$ that
we have computed exactly. We find that for all $n>0$
\begin{equation}
V_n(t)= \langle x^2(t)\rangle= \frac{v_0^2}{\Gamma^2(n+1)}\, t^{2n}\,
{}_2F_2(1,\,2 n;\,n+1,\,2 n+1;\,-2\gamma t)\, ,
\label{Vnt_exact}
\end{equation}
where $_2F_2(a_1, a_2; b_1, b_2; z)$ is the hypergeometric function~\cite{abr65,GR}.
At early times $t\ll \gamma^{-1}$, the variance grows as a power law
for any $n>0$
\begin{equation}
V_n(t) = \frac{v_0^2}{\Gamma^2(n+1)}\, t^{2n}\,
\left[1- \frac{4\, n\, \gamma}{(n+1)(2n+1)}\, t + O(t^2)\right]\, .
\label{Vnt_smallt}
\end{equation}
In contrast, the leading large $t$ behaviour depends on the value of $n$
\begin{eqnarray}
V_n(t)\approx \begin{cases}
& \frac{v_0^2}{\gamma}\, \frac{1}{(2n-1)\Gamma^2(n)}\, t^{2n-1} \, ,
\quad {\rm for} \quad n>1/2  \\
\\
& \frac{v_0^2}{\pi\,\gamma}\, \left[ \ln (2\,\gamma\, t) + \psi_0\right]\, ,
\,\,\quad {\rm for} \quad n=1/2  \\
\\
& \frac{v_0^2}{(2\gamma)^{2n}}\, \sec(n\, \pi)\, ,
\,\,\,\,\,\,\quad\quad {\rm for} \quad 0<n<1/2
\end{cases}
\label{Vnt_asymp.1}
\end{eqnarray}
where the constant $\psi_0=- \Gamma'(1/2)/\Gamma(1/2)= 1.96351\dots$.

For $n>1/2$, identifying $v_0^2/\gamma=2\,D$, the result in the first line
coincides with $\sigma_n^2(t)$ in \eqref{var_genw_n.1}, demonstrating
that the generalised RTP process \eqref{lin_rep.2} does converge
to the process \eqref{lin_rep.1} at late times. In contrast, for
$n<1/2$, the variance approaches a constant asymptotically for large $t$,
indicating that the fluctuations become time independent at late times.
The case $n=1/2$ is marginal where the variance grows very slowly
as a logarithm at late times.
Thus the exact result for the variance already
hints at a localisation transition at $n=1/2$ separating a
localised phase at late times for $n<1/2$ and a growing de-localised
phase for $n>1/2$. Extensive numerical simulations are in perfect agreement
with our results (see Section (\ref{variance})).

\vskip 0.3cm

{\noindent {\bf The position distribution $p_n(x,t)$:}}
For simplicity, we start from the initial condition
where the particle is located at the
origin $x=0$ at $t=0$, and the initial
orientation $\sigma(0)=\pm 1$ with equal probability $1/2$.
For any $n>0$, the position distribution $p_n(x,t)$ at time $t$
is symmetric in $x$ and
is supported over the interval $x\in [-x^*(t), x^*(t)]$ where
\begin{equation}
x^*(t)= \frac{v_0}{\Gamma(n+1)}\, t^n\, .
\label{x*.1}
\end{equation}
The light cone position $x^*(t)$ in \eqref{x*.1} is easy to understand.
It corresponds to the rare event when the noise $\sigma(t)$ does not change
sign up to time $t$. If it starts with $\sigma(0)=1$ (or $-1$)
and does not flip sign up to $t$, i.e., $\sigma(s)=1$ (or $-1$) for
all $0\le s\le t$,
it follows from \eqref{lin_rep.2} that this maximum
displacement of the particle
is $x^*(t)$ (or $-x^*(t)$) given in \eqref{x*.1}. The precise
form of $p_n(x,t)$ on this support depends however on the value of $n$.
We consider below the two cases $n>1/2$ and $0<n<1/2$ separately.

\begin{itemize}

\item{ {\bf The case $n>1/2$}.} In this case, the typical fluctuation
of the particle at late times is $x_{\rm typ}(t)\sim \sqrt{V_n(t)}\sim
t^{n-1/2}$
from the first line of \eqref{Vnt_asymp.1}. In contrast, the maximal
displacement on either side of the origin
(corresponding to the two edges of the support) gives another larger
scale $x^*(t)\sim t^{n}\gg x_{\rm typ}(t)$. It turns out
that the full distribution $p_n(x,t)$ has these two scales
associated to it (see Fig.~\ref{fig:schematic}).
If we look at the distribution on the
typical scale $x\sim x_{\rm typ}(t)\sim \sqrt{V_n(t)}$,
then $p_n(x,t)$ converges for large $t$ to the scaling form
\begin{equation}
p_n(x,t) \approx \frac{1}{\sqrt{V_n(t)}}\,
f\left(\frac{x}{\sqrt{V_n(t)}}\right)\, , \quad {\rm where}\quad
f(z)= \frac{1}{\sqrt{2\pi}}\, e^{-z^2/2} \, ,
\label{typ_scale.1}
\end{equation}
and $V_n(t)$ at late times is given in the first line in \eqref{Vnt_asymp.1}.
Thus as expected, the typical fluctuations are Gaussian at late times
and the scaling distribution coincides with \eqref{pnxt_intro.1} upon
identifying an effective $D= v_0^2/{2\gamma}$.

In contrast, for $|x|\sim x^*(t)\gg t^{n-1/2}$, the distribution
$p_n(x,t)$ no longer satisfies the scaling form in \eqref{typ_scale.1}.
These large values of $|x|$ represent {\em atypical} rare fluctuations
and the probability distribution
of these rare fluctuations are described by the following large deviation form
\begin{equation}
p_n(x,t)\sim \exp\left[- \gamma\, t\, \Phi_n\left(\frac{x}{x^*(t)}\right)
\right]\, ,
\label{ldev_sum.1}
\end{equation}
where we compute the rate function $\Phi_n(z)$ analytically. We find
\begin{equation}
\Phi_n(z)=  \max_{-\infty<w<\infty}\left[-z\, w + 1-
{}_2F_1\left(-\frac{1}{2},\,
\frac{1}{2\,(n-1)};\,  1+ \frac{1}{2\,(n-1)};\, -n^2\, w^2\right)\right]\, ,
\quad {\rm for} \quad n>1\, ,
\label{phinz_sum_final}
\end{equation}
where ${}_2F_1(a,b;c;z)$ is the hypergeometric function. In contrast
\begin{equation}
\Phi_n(z)=
\max_{-\infty<w<\infty}\left[-z\, w + \, 1-|w|
\,{}_2F_1\left(-\frac{1}{2},\frac{n}{2-2 n};\frac{n-2}{2 (n-1)};-\frac{1}{n^2
   |w|^2}\right)\right] \, \quad {\rm for} \quad \frac{1}{2}<n<1\, .
\label{phinzb_sum_final}
\end{equation}
The function $\Phi_n(z)$, for all $n>1/2$, is symmetric with
support in $z\in [-1,1]$ and has the small $z$ behaviour
\begin{equation}
\Phi_n(z) \simeq \frac{2n-1}{2\, n^2}\, z^2 \quad {\rm as}\,\, z\to 0\, .
\label{phinz_sum_small.1}
\end{equation}
In contrast, as $z\to 1$ (the $z\to -1$ behaviour can be obtained using
the symmetry $\Phi_n(z)=\Phi_n(-z)$), the
rate function $\Phi_n(z)$ approaches $1$ in a singular fashion
\begin{eqnarray}
\Phi_n(z) \simeq \begin{cases}
& 1 - \frac{2}{\sqrt{2\, n\, (2-n)}}\, (1-z)^{1/2}
\quad {\rm for}\quad \frac{1}{2}\le n< 2 \label{nlt1_sum.2} \\
\\
& 1- \left[\Gamma\left(\frac{2n-1}{2\,(n-1)}\right)\,
\Gamma\left(\frac{n-2}{2\, (n-1)}\right)\right]^{(n-1)/n}\,
(1-z)^{1/n} \,  \quad {\rm for}\quad n>2   \, ,  \label{ngt1_sum.2}
\end{cases}
\end{eqnarray}
with logarithmic corrections in the second term for $n=2$.
For small $z$, substituting \eqref{phinz_sum_small.1}
in \eqref{ldev_sum.1} we find that the distribution in the large deviation
regime, when extrapolated to small arguments, matches perfectly
with the Gaussian distribution describing the typical fluctuations
in \eqref{typ_scale.1}.

\item{ {\bf The case $0<n<1/2$}.} In this case, we were not able to
compute analytically the distribution $p_n(x,t)$ on the typical
scale $|x|\sim t^{n-1/2}$. However, we observed numerically that
$p_n(x,t)$ becomes time-independent for all $|x|\ll x^*(t)\sim t^n$
at late times. Moreover, this time dependent part near the
origin has a double-humped structure (see Fig.~\ref{fig:schematic}). The exact
result for the variance (in the third line of \eqref{Vnt_asymp.1})
is consistent with this time-independent form of $p_n(x,t)$, far inside
the light cones. This is the new {\em localised} phase for $0<n<1/2$,
whose origin can
be traced back to the telegraphic nature of the noise (and has no analogue
in the white noise driven process in \eqref{gen_lange.1}).
However, for large atypical fluctuations when $|x|\sim x^*(t)$
(near the edges), we show that the large deviation form
in \eqref{ldev_sum.1} continues to hold, with the rate function
given by
\begin{equation}
\Phi_n(z)=
\max_{-\infty<w<\infty}\left[-z\, w + \, 1-|w|
\,{}_2F_1\left(-\frac{1}{2},\frac{n}{2-2 n};\frac{n-2}{2 (n-1)};-\frac{1}{n^2
   |w|^2}\right)\right] \, \quad {\rm for} \quad 0<n<1/2\, .
\label{phinzc_sum_final}
\end{equation}
which is same as in \eqref{phinzb_sum_final} for $1/2<n<1$. However,
for $0<n<1/2$,
the small $z$ behaviour of $\Phi_n(z)$ in \eqref{phinzc_sum_final}
is different from that in \eqref{phinz_sum_small.1}. For $0<n<1/2$, the
leading small $z$ behaviour of $\Phi_n(z)$ is singular
\begin{equation}
\Phi_n(z) \simeq
n \left(\frac{1-n}{g(n)}\right)^{\frac{1}{n}-1}
|z|^{\frac{1}{n}}\quad {\rm as}\,\, z\to 0\, ,
\label{smallz_sum.1}
\end{equation}
where $g(n)>0$ is given by
\begin{equation}
g(n)= -\frac{n^{-\frac{n}{n-1}}  \Gamma \left(\frac{1}{2 (n-1)}\right) \Gamma
   \left(\frac{n-2}{2 (n-1)}\right)}{2 \sqrt{\pi }}\, .
\label{gn.1}
\end{equation}
For $|z|\to 1$, the behaviour of $\Phi_n(z)$ in \eqref{phinzc_sum_final}
is as in the first line of \eqref{nlt1_sum.2}.

For $|x|\ll x^*(t)\sim t^n$,
substituting the small $z$ behaviour \eqref{smallz_sum.1} in
\eqref{ldev_sum.1}, we find that the time $t$ drops out and
\begin{equation}
p_n(x,t\to \infty) \sim \exp\left[- b_n\, |x|^{1/n}\right] \, , \quad {\rm for}
\quad 1\ll |x|\ll t^n
\label{pnxb_typ.1}
\end{equation}
where the constant $b_n= \gamma\, v_0^{-1/n}\, n\,
\left((1-n)/g(n)\right)^{(1-n)/n}\, (\Gamma(n+1))^{1/n}$.
Thus, we see that at late times $p_n(x,t)$ develops
time-independent super-exponential tails (since $1/n>1$ for
$0<n<1/2$). Thus the large deviation computation also
is consistent with the observation that $p_n(x,t)$ does
become time-independent at late times for $0<n<1/2$, giving
rise to the localised phase.

In fact, this double-humped structure of the stationary
distribution for $0<n<1/2$ can be understood by studying
the limit $n\to 0$.
First, for  $t\neq s$ one has
\begin{equation}
\lim_{n\to 0}\frac{1}{\Gamma(n)}(t-s)^{n-1}= 0 ,
\label{deltarep0}
\end{equation}
since $\Gamma(n)$ has a pole at $n=0$. We then note that for $\epsilon >0 $,
\begin{equation}
\lim_{n\to 0}\int_{t-\epsilon}^t ds \frac{1}{\Gamma(n)}(t-s)^{n-1}=
\lim_{n\to 0}\frac{1}{\Gamma(n+1)}\epsilon^n
= 1 \, .
\label{deltarep2}
\end{equation}
This means that we can make the identification
$\lim_{n\to 0}\frac{1}{\Gamma(n)}(t-s)^{n-1}=\delta(t-s)$, where
$\delta(z)$ is the Dirac delta function. Thus, taking
$n\to 0$ limit in Eq. (\ref{lin_rep.2}), we see that
$x(t)= v_0\, \sigma(t)$, i.e. the process $x(t)$
is itself a telegraphic $\pm v_0$ noise.
Consequently the position distribution $p(x,t)$, assuming equal
probabilities for $\sigma(t)=\pm 1$ at equilibrium, is then exactly
given by the bimodal solution
\begin{equation}
p_0(x,t) =\frac{1}{2}\delta(x-v_0) + \frac{1}{2}\delta(x+v_0),
\label{p0exact}
\end{equation}
for all $t$. This gives the variance $V_0(t) = v_0^2$,
in agreement with the third line in Eq. (\ref{Vnt_asymp.1}) in the
limit $n\to 0$.
We thus see that the double-humped structure seen in general
for $n<1/2$ is a {\em smeared} version of what happens in
the limiting case $n\to 0$.

\end{itemize}

\vskip 0.3cm

{\noindent {\bf Exact cumulant generating function for special
values of $n$:}} We managed to compute explicitly the cumulant
generating function $U^{(n)}(t;\mu)=\langle e^{-\mu\, x(t)}\rangle$
of the process at all times $t$ for three special values of $n$,
namely $n=1$, $n=2$ and $n=1/2$. They are derived in the Appendix:
Eq. (\ref{U1sol.A1}) for $n=1$, Eq. (\ref{Ut_final.1}) for $n=2$
and Eq. (\ref{nhalf_final.1}) for $n=1/2$. While the result for $n=1$
was known earlier, the other two are new results.

\section{The trick}
\label{trick}

We consider the process \eqref{lin_rep.2}, defined for arbitrary $n>0$,
and we are interested in the one point marginal distribution of this process at
fixed $t$, i.e, the position distribution $p_n(x,t)$. In general, the
computation of this distribution is not easy for general $n>0$. In
this section, we show how it can be made easier using a simple trick. Let us
first re-write \eqref{lin_rep.2} following the change of variable
$s\to t-s$, as
\begin{equation}
x(t)= \frac{v_0}{\Gamma(n)}\, \int_0^t ds\, s^{n-1}\, \sigma(t-s)\, .
\label{lin_rep.3}
\end{equation}

Let us now define another auxiliary process ${\tilde x}(t)$ as
\begin{equation}
{\tilde x}(t)= \frac{v_0}{\Gamma(n)}\, \int_0^t ds\, s^{n-1}\, \sigma(s)\, .
\label{aux_proc.1}
\end{equation}
The process ${\tilde x}(t)$ satisfies the stochastic equation
\begin{equation}
\frac{d{\tilde x}}{dt}= \frac{v_0}{\Gamma(n)}\, t^{n-1}\, \sigma(t)\, ,
\label{aux_proc.2}
\end{equation}
which is clearly different from \eqref{dnx_tel.1} satisfied by
the original process $x(t)$. However, as we argue now, the
{\em one point marginal} distribution at fixed $t$ for the two processes
$x(t)$ and ${\tilde x}(t)$ are identical
\begin{equation}
{\rm Prob.}[x(t)=x,\, t]= {\rm Prob.}[{\tilde x}(t)=x,\, t]\, .
\label{xtx_equiv.1}
\end{equation}
This equivalence holds as long as the noise process
$\sigma(t)$ is in equilibrium.

The equivalence \eqref{xtx_equiv.1}
can be proved very simply as follows. Consider the cumulant generating
function of the process $x(t)$ in \eqref{lin_rep.3}
\begin{equation}
\langle e^{-\mu x(t)}\rangle=
\int_{-\infty}^{\infty} e^{-\mu x}\, p_n(x,t)\, dx \,
=\langle e^{-\mu\, \frac{v_0}{\Gamma(n)} \int_0^t ds\, s^{n-1}\,
\sigma(t-s)}\rangle \, .
\label{cumul_gen_x.1}
\end{equation}
Expanding the exponential in a Taylor series we get
\begin{equation}
\langle e^{-\mu x(t)}\rangle= \sum_{k=0}^\infty\frac{1}{k!}\,
\left(-\frac{\mu v_0}{\Gamma(n)}\right)^k\, \int_0^t\dots\int_0^t ds_1\dots ds_k\, s_1^{n-1}\dots s_k^{n-1}\,
\langle \sigma(t-s_1)\sigma(t-s_2)\ldots \sigma(t-s_k)\rangle\, .
\label{cumul_gen_x.2}
\end{equation}
If the noise $\sigma(t)$ is in equilibrium (which we will assume henceforth),
by definition
\begin{equation}
\langle \sigma(t-s_1)\sigma(t-s_2)\ldots \sigma(t-s_k)\rangle
= \langle \sigma(s_1)\sigma(s_2)\ldots \sigma(s_k)\rangle\ ,
\label{noise_stat.1}
\end{equation}
which follows from the time-reversal symmetry and the stationarity property
of the noise process $\sigma(t)$ when it is in equilibrium.
Substituting it back into \eqref{cumul_gen_x.2} and reconsituting the
exponential we arrive at the identity
\begin{equation}
\langle e^{-\mu x(t)}\rangle=
\langle e^{-\mu\, \frac{v_0}{\Gamma(n)} \int_0^t ds\, s^{n-1}\,
\sigma(s)}\rangle= \langle e^{-\mu\, {\tilde x}(t)}\rangle
\label{equiv.2}
\end{equation}
where we used the definition of ${\tilde x}(t)$ in \eqref{aux_proc.1}.
Since this equality holds for arbitrary $\mu$, it follows immediately that
for all $t$
\begin{equation}
p_n(x(t)=x,\, t)= p_n({\tilde x}(t)=x,\, t)\, .
\label{equiv.3}
\end{equation}
Let us remark that this equivalence between $x(t)$ and
${\tilde x}(t)$ holds even when the driving noise is white and not necessarily telegraphic.
The equivalence \eqref{equiv.3} requires that the driving noise is in equilibrium, but otherwise
holds quite generically.
Note however that this equivalence holds only
for the one point marginal distribution. The two or higher order
marginals of the two processes are evidently different. For example, even the two-time
correlation function of the two processes are not identical:
$\langle x(t_1)x(t_2)\rangle
\ne \langle {\tilde x}(t_1){\tilde x}(t_2)\rangle$.

Since in this paper we are only interested in the one point
distribution $p_n(x,t)$, we can work with the auxiliary
process ${\tilde x}(t)$ defined in \eqref{aux_proc.1} or equivalently
in \eqref{aux_proc.2} instead of the original process $x(t)$ in
\eqref{lin_rep.3}. We will see later that the process ${\tilde x}(t)$
is much simpler to study than $x(t)$. For convenience of notation,
henceforth we will refer to the process ${\tilde x}(t)$ in
\eqref{aux_proc.1} by $x(t)$.

\section{Exact result for the mean squared displacement}
\label{variance}

As a prior to computing the full distribution $p_n(x,t)$ of the
process in \eqref{aux_proc.1}, let us first
calculate its second moment (the first moment is trivially zero since $\langle \sigma(t)\rangle=0$).
Squaring \eqref{aux_proc.1} and taking average we get
\begin{equation}
V_n(t)= \langle x^2(t)\rangle= \langle {\tilde x}^2(t)\rangle=
\frac{v_0^2}{\Gamma^2(n)}\,
\int_0^t\int_0^t ds_1\, ds_2\, s_1^{n-1}\,
s_2^{n-1}\, \langle \sigma(s_1)\sigma(s_2)\rangle \, .
\label{var.1}
\end{equation}

Hence we need to compute the two-time correlation function of the
telegraphic noise $C(s_1,s_2)=\langle \sigma(s_1)\sigma(s_2)\rangle$.
Since we assume that the noise is in equilibrium, $C(s_1,s_2)= C(|s_1-s_2|)$.
The correlation function $C(s)$ can be trivially computed as follows.
Consider the product $\sigma(s_1)\sigma(s_1+s)$. This product
is $1$ if the noise has the same value at the two times $s_1$ and $s_1+s$,
otherwise it is $-1$. As we change the time from $s$ to $s+ds$,
the change in this product is either $0$ (if the noise does not flip
in $ds$) or $-2$ if the noise flips sign in the interval $ds$. Thus
\begin{eqnarray}
\sigma(s_1)\sigma(s_1+s+ds)- \sigma(s_1)\sigma(s_1+s)=
\begin{cases}
& -2 \quad {\rm with} \quad {\rm prob.}\quad \gamma\, ds \\
\\
& 0 \quad {\rm with} \quad {\rm prob.}\quad 1- \gamma\, ds \, .
\label{two_time_corr.1}
\end{cases}
\end{eqnarray}
Taking average, dividing by $ds$, followed by taking the limit
$ds\to 0$ gives
$dC(s)/ds= -2\,\gamma$. Solving, using $C(0)=1$, we get for all $s$
\begin{equation}
C(s)=\langle \sigma(s_1)\sigma(s_1+s)\rangle= e^{-2\,\gamma\, |s|} \, .
\label{corr.1}
\end{equation}

\begin{figure*}[bhtp]
    \centering
    \subfigure[\label{fig:var:simple:n2}]{
        \includegraphics[scale=1]{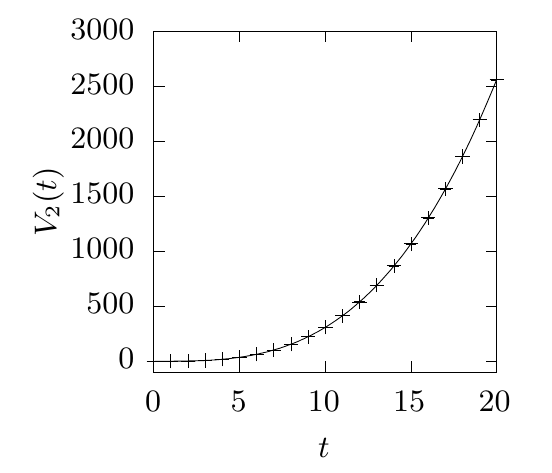}
    }
    \subfigure[\label{fig:var:simple:n05}]{
        \includegraphics[scale=1]{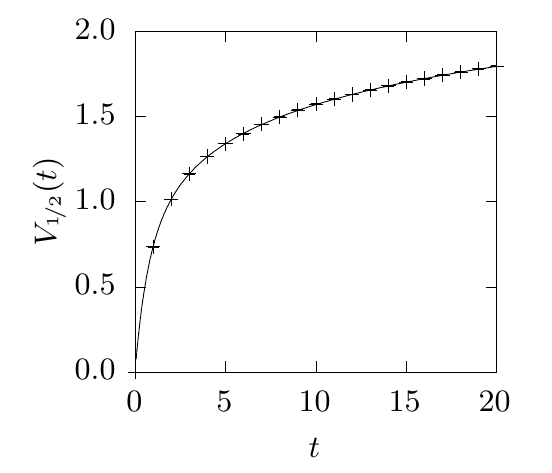}
    }
    \subfigure[\label{fig:var:simpl:n025}]{
        \includegraphics[scale=1]{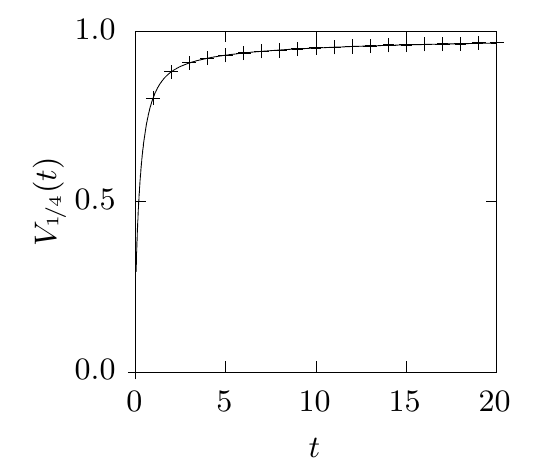}
    }
    \caption{\label{fig:variance_simple}
        The numerically obtained variance $V_n(t)$ vs.~$t$ (symbols) compared with the theoretical
        prediction \eqref{Vnt_exact} (solid curves)
        for (a) $n=2$ (b) $n=1/2$ and (c) $n=1/4$. The agreement between theory and simulation is
        perfect at all times $t$ (the numerical estimate is calculated over $10^7$ samples and their
        error bars are not visible on the scale of the symbols).
    }
\end{figure*}

Substituting \eqref{corr.1} in \eqref{var.1} gives
\begin{equation}
V_n(t)= \frac{v_0^2}{\Gamma^2(n)}\,
\int_0^t\int_0^t ds_1\, ds_2\, s_1^{n-1}\,s_2^{n-1}\,
e^{-2\, \gamma\, |s_1-s_2|}\, .
\label{var.2}
\end{equation}
Unfortunately, Mathematica was not able
to perform this double integral in \eqref{var.2} as it stands.
So, we needed to simplify a bit further. Using the symmetry of the
integrand under the exchange of $s_1$ and $s_2$, we can re-write it as
\begin{equation}
V_n(t)= \frac{2\,v_0^2}{\Gamma^2(n)}\,
\int_0^t ds_1\, s_1^{n-1} e^{-2\,\gamma\, s_1}\,
\int_0^{s_1} ds_2\, s_2^{n-1}\, e^{2\,\gamma\, s_2} \, .
\label{var.3}
\end{equation}
Next, in the integral over $s_2$, we make the rescaling
$s_2= s_1\, u$ to get
\begin{eqnarray}
V_n(t)&= & \frac{2\,v_0^2}{\Gamma^2(n)}\,
\int_0^t ds_1\, s_1^{2n-1} e^{-2\,\gamma\, s_1}\,
\int_0^{1} du\, u^{n-1}\, e^{2\,\gamma\, s_1\, u} \nonumber \\
&=& \frac{2\,v_0^2}{\Gamma^2(n)} \int_0^1 dv\, (1-v)^{n-1}\,
\int_0^{t} ds_1\, s_1^{2n-1}\, e^{-2\,\gamma\, s_1\, v}\, ,
\label{var.4}
\end{eqnarray}
where in going from the first to the second line we made a change of variable
$u=1-v$. Performing the integral over $s_1$ explicitly, we
arrive at a single integral
\begin{equation}
V_n(t)= \frac{2\, v_0^2}{(2\gamma)^{2n}\,
\Gamma^2(n)}\, \int_0^1 dv\, (1-v)^{n-1}\, v^{-2n}\,
\gamma(2\,n,\, 2\,\gamma\, v\, t)\, ,
\label{var.5}
\end{equation}
where $\gamma(a,z)= \int_0^{z} dy\, y^{a-1}\, e^{-y}$ is the incomplete
gamma function. This single integral in \eqref{var.5} can now be
done by Mathematica, leading to our explicit exact result
for the variance in \eqref{Vnt_exact} in terms of the hypergeometric function.
The asymptotic behaviour of $V_n(t)$ for small and large $t$ are
already given respectively
in \eqref{Vnt_smallt} and \eqref{Vnt_asymp.1}.
In Fig.~\ref{fig:variance_simple}, we compare our exact result
\eqref{Vnt_exact} with numerical simulations for $n=2$, $n=1/2$ and $n=1/4$, finding perfect agreement
at all times $t$.

The main consequence
of our exact result for the variance is the somewhat surprising fact
that for $n<1/2$, the variance $V_n(t)$ approaches a constant
as $t\to \infty$, leading to the emergence of a localised phase. This is
a pure consequence of the finite memory of the telegraphic noise,
and does not have any analogue when the driving noise is white, i.e.,
memoryless.

\section{Cumulant generating function via Feynman-Kac formalism}
\label{fk}

Having obtained the second moment exactly, we now turn to the full
position distribution $p_n(x,t)$. It is convenient to consider
its cumulant generating function. Noting that the noise $\sigma(t)$
at time $t$ can be either $+1$ or $-1$, we define the following
pair of cumulant generating functions
\begin{equation}
U_{\pm}^{(n)}(t;\, \mu)= \left\langle \exp\left(-\frac{\mu\, v_0}{\Gamma(n)}\,
\int_0^t ds\, s^{n-1}\, \sigma(s)\right)\, \delta_{\sigma(t),
\pm 1}\right\rangle\, .
\label{Upmt_def.n}
\end{equation}
which correspond to fixing $\sigma(t)=1$ and $\sigma(t)=-1$ respectively.
The total generating function is given by the sum
\begin{equation}
U^{(n)}(t; \mu)= U_{+}^{(n)}(t;\, \mu) + U_{-}^{(n)}(t;\, \mu) \, .
\label{Usum.n}
\end{equation}
Note that we can also write $U^{(n)}(t; \mu)$ as
\begin{equation}
U^{(n)}(t; \mu)=\left\langle e^{-\mu\, x(t)}\right\rangle=
\int_{-\infty}^{\infty} e^{-\mu\, x}\, p_n(x,t)\, dx\, .
\label{Usum_cumul.1}
\end{equation}
In fact, if we set $\mu=i\, k$, then $U^{(n)}(t; i\,k)$ is just the
Fourier transform of the position distribution at time $t$.

One can then derive the evolution equations for
$U_{\pm}^{(n)}(t;\, \mu)$ via the Feynman-Kac formalism.
Following the same route as in the case of diffusive noise (see
e.g.~\cite{maj05}),
we advance the time from $t$ and $t+dt$
and keep in mind that in time $dt$ the
noise $\sigma(t)$ flips sign with probability $\gamma\, dt$ and
stays in the same state with probability
$(1-\gamma\, dt)$. We now
split the integral inside the exponential in
\eqref{Upmt_def.n} into two parts,
one over $[0,t]$ and the other over $[t,t+dt]$.
This second part is small and we expand it
up to order $dt$. Finally taking the $dt\to 0$ limit, we get
\begin{eqnarray}
\frac{d U_+^{(n)}}{dt} &= & - \left(\gamma+ \frac{\mu\, v_0}{\Gamma(n)}\, t^{n-1}\right)\,
U_{+}^{(n)} + \gamma\, U_{-}^{(n)}\, \label{uplus_1.n} \\
\frac{d U_{-}^{(n)}}{dt}& =& - \left(\gamma-\frac{\mu\, v_0}{\Gamma(n)}\, t^{n-1}\right)\,
U_{-}^{(n)} + \gamma\, U_{+}^{(n)}\, . \label{uminus_1.n}
\end{eqnarray}
They start from the initial conditions,
\begin{equation}
U_{\pm}^{(n)}(t=0;\, \mu)=\frac{1}{2}\, .
\label{ic_1.n}
\end{equation}
These initial conditions follow from putting $t=0$
in the definition \eqref{Upmt_def.n} and assuming that at $t=0$,
$\sigma(0)$ takes values $+1$ and $-1$ with equal probability $1/2$.
Furthermore we see from the above equations that
\begin{equation}
\frac{d U_\pm^{(n)}}{dt}\Big|_{t=0} =
\mp\, \frac{1}{2}\frac{\mu\, v_0}{\Gamma(n)}\ t^{n-1}\Big|_{t=0}.
\label{init_cond.1}
\end{equation}
Thus at $t=0$, while the derivatives vanish for $n>1$, they diverge
for $n<1$.

We now rewrite the pair of equations \eqref{uplus_1.n} and \eqref{uminus_1.n}
in terms of $U^{(n)}(t; \mu)= U_{+}^{(n)}(t;\, \mu) + U_{-}^{(n)}(t;\, \mu) $ and
$W^{(n)}(t; \mu)= U_{+}^{(n)}(t;\, \mu) - U_{-}^{(n)}(t;\, \mu) $ which yields
\begin{eqnarray}
\frac{dU^{(n)}}{dt}&=& -\frac{\mu v_0t^{n-1}}{\Gamma(n)}\, W^{(n)}\label{u1}\\
\frac{dW^{(n)}}{dt} &=& -2\,\gamma\, W^{(n)}
-\frac{\mu v_0 t^{n-1}}{\Gamma(n)}\, U^{(n)} \, . \label{u2}
\end{eqnarray}
They satisfy the initial conditions
\begin{equation}
U^{(n)}(t=0;\mu)=1 \, , \quad {\rm and}\quad W^{(n)}(t=0;\mu)=0\, .
\label{uw_init.1}
\end{equation}
Eliminating $W^{(n)}$ by using Eq.~(\ref{u1}) in Eq.~(\ref{u2}), yields
a closed second order differential equation for $U^{(n)}$
\begin{equation}
\frac{d^2 U^{(n)}}{dt^2} +\left(2\gamma- \frac{(n-1)}{t}\right)\,
\frac{d U^{(n)}}{dt} -\frac{\mu^2\, v_0^2\, t^{2n-2}}{\Gamma^2(n)}\,
U^{(n)}=0 \, ,\label{ueq}
\end{equation}
valid for all $t\ge 0$ and all $n>0$.
To solve this second order equation, we need
two boundary conditions that read
\begin{equation}
U^{(n)}(t=0;\mu)=1 \, \quad {\rm and}\quad \frac{dU^{(n)}}{dt}\Big|_{t=0}=
\frac{\mu^2\, v_0^2}{n\, \Gamma^2(n)}\, t^{2n-1}\Big|_{t=0}\, .
\label{un_bc.1}
\end{equation}
The second condition can be derived by using \eqref{uw_init.1}
in \eqref{u1} and \eqref{u2} as follows. Substituting $U^{(n)}(t=0;\mu)=1$ on
the right hand side of \eqref{u2}, and solving for $W^{(n)}$ at short times
gives to leading order, $W^{(n)}(t;\mu)\approx - (\mu\, v_0/\Gamma(n+1))\, t^n$.
Substituting this in \eqref{u1} yields the second condition in \eqref{un_bc.1}.
Thus, for $n>1/2$, the derivative of $U^{(n)}$ vanish at $t=0$, while
for $n<1/2$ it diverges. Exactly at $n=1/2$, the derivative at $t=0$
is a constant and the initial conditions read
\begin{equation}
U^{(1/2)}(t=0;\mu)=1 \, \quad {\rm and}\quad \frac{dU^{(1/2)}}{dt}\Big|_{t=0}=
\frac{2\,\mu^2\, v_0^2}{\pi}\, .
\label{unhalf_bc.1}
\end{equation}

There is an alternative way to arrive at a closed second order differential equation for
$U^{(n)}_{+}(t;\mu)$ and $U^{(n)}_{-}(t;\mu)$ separately. This second representation turns out to
be useful also for deriving the exact result for some values of $n$, such as $n=2$ (as shown
in the Appendix).
To proceed further, it is useful first to write down the pair of equations
\eqref{uplus_1.n} and \eqref{uminus_1.n} in an operator form as follows
\begin{eqnarray}
\hat{L}_+\, U_{+}^{(n)} & = & \gamma\, U_{-}^{(n)}\,; \quad {\rm with}\quad
\hat{L}_+= \frac{d}{dt}+ \gamma+ \frac{\mu\, v_0}{\Gamma(n)}\, t^{n-1} \label{L+.n} \\
\hat{L}_-\, U_{-}^{(n)} & = & \gamma\, U_{+}^{(n)}\,; \quad {\rm with}\quad
\hat{L}_-= \frac{d}{dt}+ \gamma- \frac{\mu\, v_0}{\Gamma(n)}\, t^{n-1} \, . \label{L-.n}
\end{eqnarray}
We then operate the first equation from the left by $\hat{L}_{-}$ and use the second equation to
write a closed equation for $U_+^{(n)}(t;\, \mu)$ only. Similarly, one can obtain a
closed equation for
$U_{-}^{(n)}(t;\, \mu)$ also. We get
\begin{eqnarray}
\hat{L}_{-}\, \hat{L}_+ U_+^{(n)} &= & \gamma^2\, U_+^{(n)} \, \label{U+t.n} \\
\hat{L}_{+}\, \hat{L}_- U_-^{(n)} &= & \gamma^2\, U_-^{(n)}\, . \label{U-t.n}
\end{eqnarray}
Expanding the operators, we get a pair of ordinary differential equations for
$U_{\pm}^{(n)}(t;\, \mu)$
\begin{eqnarray}
\frac{d^2U_+^{(n)}}{dt^2}+ 2\, \gamma\, \frac{dU_+^{(n)}}{dt}+
\left(\frac{\mu\, v_0}{\Gamma(n-1)}\, t^{n-2}- \frac{\mu^2\, v_0^2}{\Gamma^2(n)}\, t^{2n-2}\right)\,
U_+^{(n)} &= & 0 \label{U+only.n} \\
\frac{d^2U_-^{(n)}}{dt^2}+ 2\, \gamma\, \frac{dU_-^{(n)}}{dt}+
\left(-\frac{\mu\, v_0}{\Gamma(n-1)}\, t^{n-2}- \frac{\mu^2\, v_0^2}{\Gamma^2(n)}\, t^{2n-2}\right)\,
U_-^{(n)}& =& 0 \, , \label{U-only.n}
\end{eqnarray}
These equations have to be solved with the
initial conditions \eqref{ic_1.n} and \eqref{init_cond.1}.
It is also clear that the solutions satisfy the
following symmetry
\begin{equation}
U_{+}^{(n)}(t;\, \mu)= U_{-}^{(n)}(t;\, -\mu)\, .
\label{symmetry.n}
\end{equation}
Note that in this representation one first solves for
$U^{(n)}_+(t;\mu)$ and $U^{(n)}_{-}(t; \mu)$
separately from Eqs. (\ref{U+only.n}) and (\ref{U-only.n}) and
then adds up these solutions to compute $U^{(n)}(t;\mu)$.

For certain specific values of $n$ such as $n=1$, $n=2$ and $n=1/2$, we can solve
these differential equations explicitly, as shown in the Appendix. For example, it turns out that for $n=1$
and $n=1/2$ one can solve directly the differential equation \eqref{ueq} for $U^{(n)}(t;\mu)$.
In contrast, for $n=2$, it turns out to be more convenient to use the second path, i.e,
first solve Eqs. (\ref{U+only.n}) and (\ref{U-only.n}) separately and then add them up.
These exact results are presented in the Appendix.
For generic $n$, finding an explicit solution valid at all $t$
seems difficult. However, for large $t$, one can make progress
as we demonstrate in the next section. In particular, for large $t$ and
large $x$, keeping $x/x^*(t)$ fixed (where we recall $x^*(t)=
v_0\, t^n/\Gamma(n+1)$), we show below how the large deviation
properties of $p_n(x,t)$ can be extracted from Eq. (\ref{ueq}).

\section{Large deviation analysis of the cumulant generating function}
\label{ld}

It is easier and perhaps more physical to guess the possible large deviation behaviour
of $p_n(x,t)$ in real space $x$, rather than for its cumulant generating function
$U^{(n)}(t;\mu)$. So, our strategy would be to (i) first guess the large
deviation form of $p_n(x,t)$ and then use it to anticipate the
large deviation form of $U^{(n)}(t;\mu)$ and then (ii) substitute
this anticipated form in the differential equation \eqref{ueq}
to explicitly derive the large deviation function.

\vskip 0.3cm

{\noindent {\bf Anticipated large deviation form for $p_n(x,t)$.}}
Let us then first see
what we may expect for the large deviation behaviour of
$p_n(x,t)$. It is clear that if the noise
$v_0\, \sigma(t)$ does not flip sign at all in time $t$,
then the maximum distance travelled
by the particle is $\pm v_0\, t^n/\Gamma(n+1)$ from \eqref{aux_proc.1}.
This is the largest possible deviation in the $x$ direction, and the
probability for this event (of no flipping) is
clearly $e^{-\gamma\,t}$.  Hence, it is natural to
anticipate that
in the limit $t\to \infty$, $x\to \infty$ but with the
ratio $z=\Gamma(n+1)\,x/(v_0\,t^n)$ fixed, the
distribution $p_n(x,t)$ exhibits the following large deviation behaviour
\begin{equation}
p_n(x,t) \sim \exp\left[- \gamma\, t\,
\Phi_n\left(\frac{\Gamma(n+1)\, x}{v_0\, t^n}\right)\right]\, ,
\label{ldp.n}
\end{equation}
where $\Phi_n(z)$ is the rate function, which is symmetric in $z$
and is supported
over the interval $z\in [-1,1]$ since $|z|$ can not exceed $1$.
The probability
of the rarest event that $|z|=1$ is $e^{-\gamma \, t}$, i.e.,
the probability that $\sigma(t)$
does not change sign up to time $t$.
Hence, putting $z=1$ in \eqref{ldp.n}, we
also infer that
\begin{equation}
\Phi_n(\pm 1)= 1\, .
\label{phi1.n}
\end{equation}

\vskip 0.3cm

{\noindent {\bf Anticipated large deviation form for $U^{(n)}(t;\mu)$.}}
Let us now see what Eq. \eqref{ldp.n} would imply for the
cumulant generating function $U^{(n)}(t;\mu)$ in \eqref{Usum_cumul.1}.
Substituting the anticipated large deviation behaviour
\eqref{ldp.n} in \eqref{Usum_cumul.1} we get
\begin{eqnarray}
U^{(n)}(t;\, \mu)&= & \int_{-\infty}^{\infty} dx\,
p_n(x,t)\, e^{-\mu\, x} \nonumber \\
& \sim & \int_{-1}^{1} dz\, e^{- \gamma\, t\,
\Phi_n(z)-\frac{\mu\, v_0\, t^n}{\Gamma(n+1)}
\, z} \nonumber \\
&\sim &  \int_{-1}^{1}dz\, e^{- \gamma\, t\,
\left[ w\, z + \Phi_n(z)\right]}\, ; \quad {\rm with}\,\,
w=\frac{\mu\, v_0\, t^{n-1}}{\Gamma(n+1)\, \gamma}
\label{Ut_ldp.n}
\end{eqnarray}
where, in going from the first to the second line, we made the
change of variable
$x= (v_0\, t^n/\Gamma(n+1))\, z$, and we did not keep track of
pre-exponential factors.

We now take the limit $t\to \infty$ and $\mu\to 0$ for $n>1$
(or $\mu\to \infty$ for $n<1$), such that
$w= \mu\, v_0\, t^{n-1}/(\Gamma(n+1)\,\gamma)$ remains fixed.
Then, for large $t$, we can estimate the
integral in \eqref{Ut_ldp.n} by the saddle point method. This gives
\begin{equation}
U^{(n)}(t;\, \mu) \sim e^{- \gamma\, t\, H_n(w)}\, ; \quad {\rm where}\quad
H_n(w)= \min_{-1\le z\le 1}\left[ w\, z+ \Phi_n(z)\right]\, .
\label{Ut_ldp.n2}
\end{equation}
Hence, we have our desired large deviation ansatz for $U^{(n)}(t; \mu)$
\begin{equation}
U^{(n)}(t;\, \mu) \sim \exp\left[- \gamma\, t\,
H_n\left(\frac{\mu\, v_0\, t^{n-1}}{\Gamma(n+1)\, \gamma}\right)\right] \, .
\label{uplus_ldp.n1}
\end{equation}

Let us now check that this large deviation ansatz \eqref{uplus_ldp.n1}
is consistent with the extreme trajectories.
Consider the limit $\mu\to \infty$ first. In this limit,
from \eqref{Upmt_def.n}, it follows that
the extreme paths starting with
$\sigma=-1$ that never flip will dominate and contribute
\begin{equation}
U^{(n)}(t;\, \mu)\sim
\exp\left(-\gamma t +
\mu \frac{v_0\, t^n}{\Gamma(n+1)}\right) \, . \label{as1}
\end{equation}
Similarly for $\mu\to-\infty$ we find
\begin{equation}
U^{(n)}(t;\, \mu)\sim \frac{1}{2}\exp\left(-\gamma t -
\mu \frac{v_0\, t^n}{\Gamma(n+1)}\right)\, .\label{as2}
\end{equation}
We note that both \eqref{as1} and \eqref{as2} do satisfy
the large deviation ansatz in \eqref{uplus_ldp.n1} with
a predicted asymptotic behaviour of $H_n(w)$ for large $|w|$
\begin{equation}
H_n(w) \approx 1 - |w|
\label{hinf}
\end{equation}

\vskip 0.3cm

{\noindent {\bf Explicit solution for $H_n(w)$.}} Thus our main conclusion
from the above exercise is that the cumulant generating function
$U^{(n)}(t; \mu)$ for large $t$ has an anticipated large deviation \eqref{uplus_ldp.n1} with
$H_n(w)$ denoting the rate function in the $w$ space. To derive
$H_n(w)$ explicitly, we substitute
the ansatz \eqref{uplus_ldp.n1}
in the differential equation \eqref{ueq} and obtain
\begin{equation}
\gamma^2\, \left[ G_n^2(w)-2\, G_n(w)-n^2\, w^2\right] +
\frac{1}{t}\,\gamma\, (n-1)\,
\,\left[G_n(w)-w G_n'(w)\right]=0\, .
\label{Hnw.1}
\end{equation}
where
\begin{equation}
G_n(w)= H_n(w)+ (n-1)\,w\, H_n'(w)\, .
\label{Gn_def.1}
\end{equation}
Consequently, the leading order term for large $t$ gives
\begin{equation}
G_n^2(w)-2\, G_n(w)-n^2\, w^2=0\, .
\label{Gnw.1}
\end{equation}
Solving the quadratic equation gives a first order differential equation for $H_n(w)$
\begin{equation}
(n-1)\, w\, H_n'(w)+ H_n(w)= 1\pm  \sqrt{1+ n^2\, w^2}\, \label{Hnw.2}
\end{equation}
where we have the choice of two roots. We now consider
the two cases $n> 1$ and $0<n< 1$ separately, as it will turn out
that the solution $H_n(w)$ has different forms in these two cases.

\begin{itemize}

\item{{\bf The case $n> 1$.}} In this case, as $t\to \infty$, in order to keep
$w= \mu\, v_0\, t^{n-1}/{\gamma\Gamma(n+1)}$ fixed, we must have
$\mu\to 0$.
Solving the first-order equation \eqref{Hnw.2}
gives the general solution
\begin{equation}
H_n(w) =  \frac{w^{-1/(n-1)}}{n-1}\left[c+ \int_0^w
\left(1\pm \sqrt{1+n^2\, x^2}\right)\, x^{(2-n)/(n-1)}\, dx \right],
\label{Hnw_first}
\end{equation}
where $c$ is an integration constant.
To determine $c$ and the sign of the root to be
chosen we note that we must have $H_n(w\to 0)=0$,
which follows from the fact that $U^{(n)}(t;\, \mu=0)=1$ for all $t$.
However by symmetry of the probability distribution
we also have $H'_n(0)=0$. Examining Eq.~(\ref{Hnw.2})
then shows that we should choose the negative root.
We thus find
\begin{equation}
H_n(w) =  \frac{w^{-1/(n-1)}}{n-1}\left[c+ \int_0^w \left(1-
\sqrt{1+n^2\, x^2}\right)\, x^{(2-n)/(n-1)}\, dx \right],
\label{hint}
\end{equation}
and with this choice of sign we see that the integrand in
\eqref{hint} for small $x$  behaves as $x^{\frac{n}{(n-1)}}$ and the integral is
thus convergent around $x=0$ if $n> 1$.
For small $w$ we thus find
\begin{equation}
H_n(w) = c\frac{w^{-1/(n-1)}}{n-1} -\frac{n^2}{2(2n-1)} w^2,
\end{equation}
and from this we conclude that $c=0$ since $H_n(w\to 0)=0$.
Setting $c=0$ and performing the integral explicitly in
\eqref{hint} we get for $n>1$
\begin{equation}
H_n(w)= 1- {}_2F_1\left(-\frac{1}{2},\, \frac{1}{2\,(n-1)};\,
1+ \frac{1}{2\,(n-1)};\, -n^2\, w^2\right),
\label{Hnw_sol.1}
\end{equation}
where ${}_2F_1(a,b;c;z)$ is the hypergeometric function~\cite{GR}.
For $n=1$ and $n=2$ one obtains rather simple expressions
\begin{eqnarray}
H_1(w) & = & 1-\sqrt{1+w^2} \label{hnw1} \\
H_2(w) &= & 1- \frac{1}{2}\, \sqrt{1+ 4\, w^2}- \frac{{\rm arcsinh}(2\, w)}{4\, w}\, .
\label{hnw2}
\end{eqnarray}

The function $H_n(w)$ in \eqref{Hnw_sol.1} is clearly symmetric around $w=0$, i.e., $H_n(w)=H_n(-w)$. The small $w$
asymptotic behaviour of $H_n(w)$ is easy to derive and we get for all $n\ge 1$
\begin{equation}
H_n(w)= -\frac{n^2}{2\,(2n-1)}\, w^2+ \frac{n^4}{8\,(4n-3)}\, w^4 -\frac{n^6}{16\,(6n-5)}\, w^6 +
O\left(w^8\right) \quad {\rm as}\quad w\to 0 \, . \label{Hnw_small.1}
\end{equation}
In contrast, the large $w$ asymptotics of
$H_n(w)$ depends on whether $n<2$ or $n>2$. We get as $w\to \infty$
\begin{eqnarray}
H_n(w)= \begin{cases}
& -w +1 - \frac{1}{2\,n\,(2-n)}\, \frac{1}{w} + \ldots \quad
{\rm for} \quad 1\le n<2 \label {nlt1.1} \\
\\
& -w + 1 - (n-1)\, \Gamma\left(\frac{2n-1}{2\,(n-1)}\right)\,
\Gamma\left( \frac{n-2}{2\, (n-1)}\right)\, n^{-n/(n-1)}\, w^{-1/(n-1)}+ \ldots
{\rm for} \quad  n>2 \label {ngt1.1}\, .
\end{cases}
\end{eqnarray}

\item{{\bf The case $0<n<1$.}}
In this case if we consider the
limit where $t\to\infty$ while
$w=\mu\, v_0\, t^{n-1}/(\Gamma(n+1)\, \gamma)$ is held fixed,
this implies that we are considering the limit
where $\mu\to \infty$. Thus even when $w$ is small one is still
in the limit where $\mu\to\infty$. This means that we
can no longer use the boundary condition at $\mu=0$ or $w=0$ as
in the case $n>1$. Instead we have to use the asymptotic boundary
condition in Eq.~(\ref{hinf}) as $|w|\to\infty $
to determine the behaviour of $H_n(w)$ for $w$
finite (precisely the opposite of what we have done for $n>1$).
now, the general solution of the first-order differential
equation \eqref{Hnw.2} can also be written as
\begin{equation}
H_n(w) =  \frac{-w^{-1/(n-1)}}{n-1}\left[c_1+ \int_w^{\infty}
\left(1\pm \sqrt{1+n^2\, x^2}\right)\, x^{(2-n)/(n-1)}\, dx \right],
\label{Hnw_second}
\end{equation}
where $c_1$ is the integration constant.
next, we note that for large $w$ the
indefinite form of the integral appearing in Eq.~(\ref{Hnw_second})
\begin{equation}
I_\pm(w) = \int^w dx
\left(1\pm \sqrt{1+n^2\, x^2}\right)\,
x^{(2-n)/(n-1)} \approx (n-1)\left[w^{\frac{1}{(n-1)}}
\pm w^{\frac{n}{(n-1)}}\right],
\label{Ipmw.1}
\end{equation}
is convergent as $w\to \infty$ for $0<n<1$. Now,
in order to satisfy the boundary condition in Eq.~(\ref{hinf}),
we see using \eqref{Ipmw.1} in \eqref{Hnw_second} that
we must choose again the negative root above and set
$c_1=0$ in Eq. (\ref{Hnw_second}). This gives
\begin{equation}
H_n(w) = - \frac{w^{-1/(n-1)}}{n-1}\left[\int_w^\infty \left(1- \sqrt{1+n^2\, x^2}\right)\, x^{(2-n)/(n-1)}\, dx \right].
\end{equation}
The integral above can be evaluated to yield (for $w>0$)
\begin{equation}
H_n(w)=1-w\, {}_2F_1\left(-\frac{1}{2},\frac{n}{2-2 n};\frac{n-2}{2 (n-1)};-\frac{1}{n^2
   w^2}\right).
\end{equation}
The solution for $w<0$ can be be similarly obtained. In fact, since $H_n(w)$ is symmetric, the solution
for all $w$ is given by
\begin{equation}
H_n(w)=1-|w|\, {}_2F_1\left(-\frac{1}{2},\, \frac{n}{2-2 n};\, \frac{n-2}{2 (n-1)};\, -\frac{1}{n^2\,w^2}\right)\, .
\label{Hnw_nlt1}
\end{equation}

This solution has the large $w$ expansion
\begin{equation}
H_n(w)= 1-|w| -\frac{1}{2 n(2-n)  |w|}+\cdots ,
\end{equation}
that coincides with the large $w$ expansion for the case $1\leq n <2$ given in Eq.~(\ref{nlt1.1}).
The small $w$ expansion is given by
\begin{equation}
H_n(w)= \frac{n^{-\frac{n}{n-1}}  \Gamma \left(\frac{1}{2 (n-1)}\right) \Gamma
   \left(\frac{n-2}{2 (n-1)}\right)}{2 \sqrt{\pi }}|w|^{\frac{1}{1-n}}-\frac{n^2}{2\,(2n-1)}\, w^2+ \frac{n^4}{8\,(4n-3)}\, w^4 -\frac{n^6}{16\,(6n-5)}\, w^6 +
O\left(w^8\right).
\end{equation}
This has the same analytic terms as the expansion Eq.~(\ref{Hnw_small.1}) which is valid for the case $n>1$ but
the first term is non-analytic, proportional to $|w|^{\frac{1}{1-n}}$ which appears only for $0<n<1$.
In the case $n>1/2$, the leading correction to $H_n(w)$ at small $w$ is still of order $w^2$,
however for $n<1/2$ it is the term proportional to $|w|^{\frac{1}{1-n}}$.
Therefore at lowest order for small $w$ we find
\begin{eqnarray}
H_n(w)= \begin{cases}
& -\frac{n^2}{2\,(2n-1)}\, w^2+ \ldots \quad
{\rm for} \quad \frac{1}{2}<n \leq 1 \label {nlt1_new.1} \\
& -g(n)|w|^{\frac{1}{1-n}}+ \ldots
\quad {\rm for} \quad 0<n <\frac{1}{2} .\label {ngt1_new.1}\,
\end{cases}
\end{eqnarray}
where
\begin{equation}
g(n)= -\frac{n^{-\frac{n}{n-1}}  \Gamma \left(\frac{1}{2 (n-1)}\right) \Gamma
   \left(\frac{n-2}{2 (n-1)}\right)}{2 \sqrt{\pi }},
\end{equation}
and one can check that $g(n) >0$ for $0<n <\frac{1}{2}$.

Finally, we note that exactly at $n=1/2$, \eqref{Hnw_nlt1} has a
simpler expression
\begin{equation}
H_{1/2}(w)= 1- \frac{1}{2}\,\sqrt{w^2+4}-\frac{w^2}{4}\, {\rm arcsinh}\left(\frac{2}{|w|}\right)\, .
\label{h_halfw.1}
\end{equation}
For small $w$, we get $H_{1/2}(w)\approx - w^2\, \ln|w|/4$ to leading order, while for large $|w|$,
$H_{1/2}(w) \approx 1- |w| $.

\end{itemize}

\section{Extracting the large deviation behaviour of $p_n(x,t)$}\label{inv}

\begin{figure}
\includegraphics[width=0.7\textwidth]{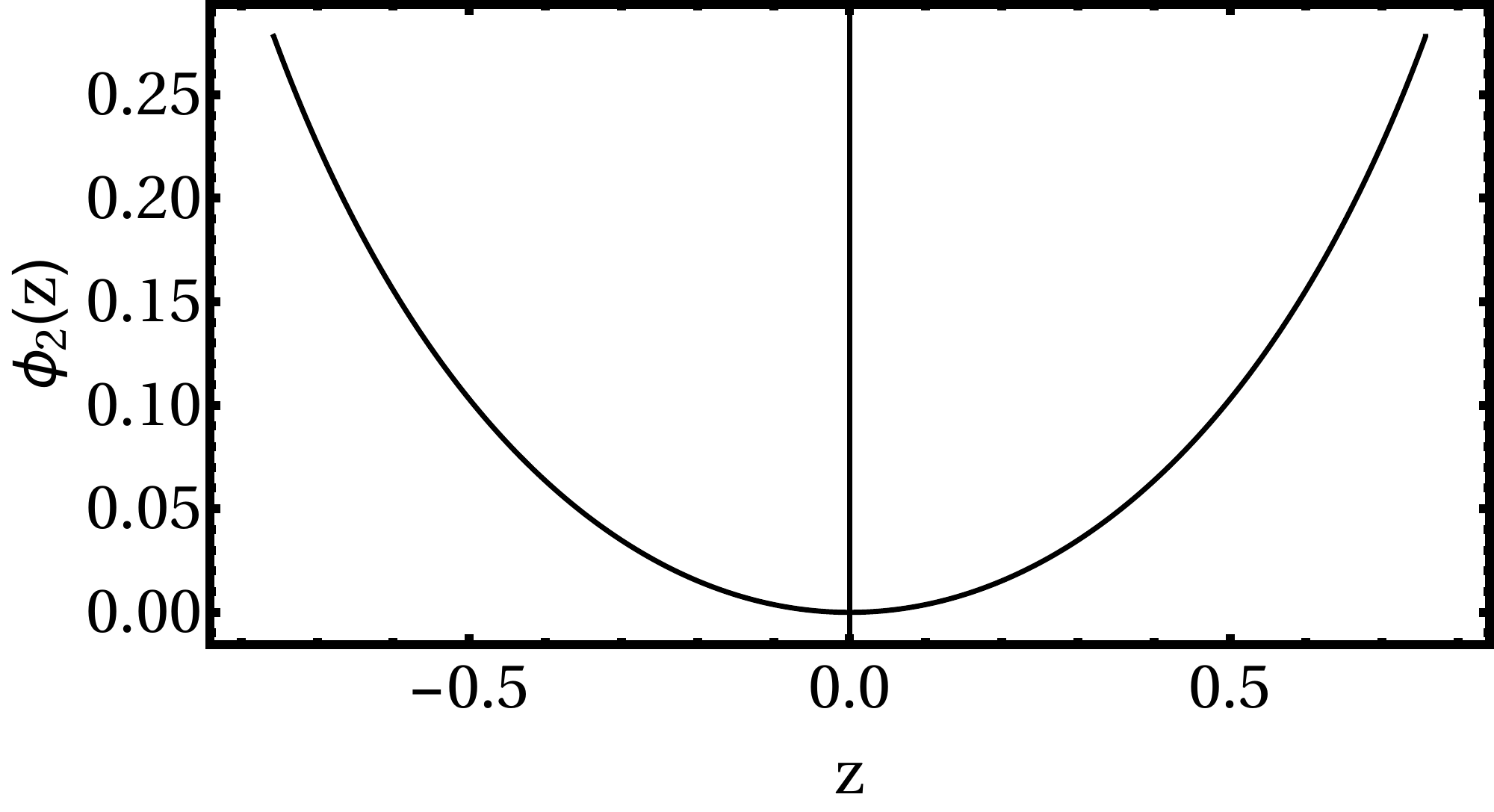}
\caption{The rate function $\Phi_2(z)$ vs. $z$ obtained parametrically from
Eq. (\ref{phiw.1}) for $n=2$. The function $\Phi_2(z)$ is symmetric around $z=0$
and is supported over $z\in [-1,1]$. }
\label{fig:Phi2z}
\end{figure}

We have already seen that on the scale $x\sim x^*(t)= v_0 t^n/\Gamma(n+1)$, the
distribution $p_n(x,t)$ has the large deviation form in \eqref{ldp.n} where
the rate function $\Phi_n(z)$ is related to $H_n(w)$ via the Legendre transform
in \eqref{Ut_ldp.n2}. Formally inverting this Legendre transform we get
\begin{equation}
\Phi_n(z)= \max_{-\infty<w<\infty}\left[-z\, w + H_n(w)\right]\, ,
\label{phiz.n}
\end{equation}
where $H_n(w)$ is known explicitly from \eqref{Hnw_sol.1} for $n>1$ and
\eqref{Hnw_nlt1} for $0<n<1$. As an example, consider first $n=1$
for which $H_1(w)=1-\sqrt{1+w^2}$ from \eqref{hnw1}. Maximizing \eqref{phiz.n} gives
\begin{equation}
\Phi_1(z)= 1- \sqrt{1-z^2} \, \quad -1\le z\le 1\, ,
\label{Phi1z.inv}
\end{equation}
which thus reproduces the result \eqref{phi1z_intro.1} quoted in the introduction.
For generic $n$, it is difficult to obtain $\Phi_n(z)$ explicitly.
However, by maximising the function inside
the parenthesis in \eqref{phiz.n} with respect to $w$, we can express
$\Phi_n(z)$ in the following parametric form that can be easily
plotted in Mathematica
\begin{eqnarray}
z &= & H_n'(w) \nonumber \\
\Phi_n & = & - w\, H_n'(w)+H_n(w)
\label{phiw.1} .
\end{eqnarray}
As an example, consider $n=2$ for which $H_2(w)$ is given
explicitly in \eqref{hnw2}. Using this in
\eqref{phiw.1}, we plot $\Phi_2(z)$ vs. $z$ in Fig.~\ref{fig:Phi2z}.

The rate function is symmetrically supported over the interval $z\in [-1,1]$.
The asymptotic behaviors of $\Phi_n(z)$ as $z\to 0$ and $z\to \pm 1$ can be
obtained from \eqref{phiz.n} by substituting the $w\to 0$
and $w\to \mp \infty$ behaviors of $H_n(w)$ detailed in the previous section.
This leads to the results quoted in the summary in Section \ref{summary}.

To verify our analytical prediction for $\Phi_n(z)$ in \eqref{phiz.n} we have
done extensive simulations using a nontrivial importance sampling algorithm
that is described in detail in the next section. The simulation results are
in excellent agreement with our analytical predictions.

\section{Numerical simulations}\label{numerics}

Here we compare the analytical results of the previous sections with numerical simulations. We first describe
the method of simulation and in particular the crucial idea of importance sampling. We then carry out the
comparison between numerics and analytics.

\subsection{The Method}

Performing computer simulations to study any model, requires a suitable
discretisation of the model. Fortunately in this case, an exact
discretisation is possible. We use for all values of $n$
the representation \eqref{aux_proc.1} which can be
exactly discretised by a piecewise solution of the integral
\begin{align}
    x(t) = \frac{v_0}{\Gamma(n+1)} \sum_i \sigma_i (t_{i+1}^n - t_i^n),
\end{align}
where the sum goes over all flips, such that the $\sigma_i$ alternate between $+1$ and $-1$.

Now we can generate independent samples of this process for arbitrary $t$ by
drawing the times until the next flip $\tau_i$ from an exponential
distribution $P(\tau) = \gamma\, \exp(-\gamma\,\tau)$
until $\sum_i \tau_i \ge t$.
The last waiting time is truncated to enforce $\sum_i \tau_i = t$.
Note that the number of flips fluctuates.

To obtain numerical estimates for the
rate function, it is necessary to calculate the probability density function,
especially including the far tails of extremely rare events which occur with
probabilities of less than, say, $10^{-100}$. It is infeasible with
current computers to generate  the order of $10^{100}$ samples, which are necessary
to observe such an event once on average---much less to generate enough of
those rare events to allow an estimate of their probability with reasonable
statistical precision.

Therefore, we need to resort to more sophisticated Markov chain Monte Carlo
simulation techniques. The basic idea is to generate samples which are biased
in a controlled way to increase the chance to encounter a very rare event, thus
allowing to collect robust statistics of this event. Since the bias is well
controlled, one can obtain the unbiased probability density from these measurements.
This fundamental concept is also known as \emph{importance sampling}.

To generate samples with a well controlled bias, we use the \emph{Metropolis-Hastings}
algorithm \cite{Hastings1970monte}. Therefore we construct a Markov chain of configurations, in our case
the vector of times between flips $\vec\tau_i$. To generate the next link $\vec\tau_{i+1}$
in the Markov chain, we propose a configuration $\vec\tau'$ generated by applying
a small change to the current configuration $\vec\tau_i$
and accept it according to an acceptance probability $p_\mathrm{acc}(\vec\tau_i \to \vec\tau')$.
If the new configuration is accepted $\vec\tau_{i+1} = \vec\tau'$, otherwise the
old configuration is repeated in the chain, i.e.\ $\vec\tau_{i+1} = \vec\tau_i$.
The acceptance probability needs to be chosen such that detailed balance holds.
The exact choice then determines according to which distribution the configurations
will appear eventually in the Markov chain. The change move in our case is to select
a random component $\tau_i^*$ and replacing it with a new random time drawn from
the same exponential distribution, or flip the initial direction $\sigma_0$.
Note that this might change the number of $\tau_i$ defining the configuration: if
the newly generated $\tau_i^*$ is larger, the last few $\tau_i$ might need to be removed
and if $\tau_i^*$ becomes smaller, a few more flips might need to happen before
the time $t$ is reached.
Also this change move allows to reach every possible configuration
after enough changes, which means that ergodicity holds, the second prerequisite
necessary for a Markov chain to generate configurations according to the desired distribution.

Here we use for the acceptance probability the original choice generating Boltzmann
distributed states $p_\mathrm{acc} = \min\left(1, e^{-\Delta x/T}\right)$ \cite{metropolis1953equation},
i.e.\ configurations in the Markov chain will be distributed according to
\begin{align}
    \label{eq:biased_dist}
    Q_T(\vec\tau) = \frac{1}{Z_T} e^{-x(\vec\tau)/T} Q(\vec\tau),
\end{align}
where $Q$ is the natural distribution and $Z_T$ the partition function
necessary for normalization. The ``temperature''
$T$ is in this context just a free parameter, which we can use to bias the
resulting samples: small temperatures will lead to small ``energies'' $x$,
large $T$ lead to typical values of $x$ and small negative $T$ lead to large
values of $x$.

We can estimate the probability density $p_T(x)$ of our artificial temperature
ensemble and remove the bias to get the unbiased distribution in a range of very
atypical $x$ but with good statistics \cite{Hartmann2011}. Using Eq.~(\ref{eq:biased_dist}) we see
\begin{align}
    p_T(x) &= \sum_{\{\vec\tau | x(\vec\tau) = x \}} Q_T(\vec\tau)\\
            &= \sum_{\{\vec\tau | x(\vec\tau) = x \}} \frac{1}{Z_T} e^{-x(\vec\tau)/T} Q(\vec\tau)\\
            &= \frac{1}{Z_T} e^{-x/T} p(x),
\end{align}
where $p(x)$ is the searched for, unbiased distribution.
This means that we need the partition function $Z_T$ to correct the bias. Fortunately,
we can use that $p(x)$ needs to be unique if derived for the same $x$ but different $T$.
So we need to simulate and estimate $p_T(x)$ for multiple, carefully chosen values of $T$,
such that the different $p_T(x)$ overlap. Then the ratio of the $Z_T$ of overlapping
ranges can be determined using
\begin{align}
    p_{T_j}(x) e^{x/T_j} Z_{T_j} = p_{T_i}(x) e^{x/T_i} Z_{T_i}.
\end{align}
The absolute values of the $Z_T$ are then obtained by normalization of the full distribution
$p(x)$.

As usual for Markov chain Monte Carlo, care has to be taken that the Markov chain is
equilibrated before taking measurements and the correlation between two samples has to be considered to
avoid underestimation of the statistical error \cite{newman1999monte}.
Generally, this method works quite well for a wide range
of problems \cite{schawe2018avoiding,hartmann2018distribution,Schawe2018,borjes2019large}, especially it was already applied in the context of the run-and-tumble
particle, the $n=1$ case of the model at hand \cite{Hartmann2020}.

\subsection{Comparison of analytical and numerical results}
    First, we generated $10^7$ independent trajectories and measured
    their positions at time $t$ to obtain
    estimates for the variance $V_n(t)=\langle x^2(t)\rangle$
    shown in Fig.~\ref{fig:variance_simple} as well as an
    estimate for the distribution $p_n(x, t)$ shown in Fig.\ref{fig:schematic}.

\begin{figure}
        \centering
        \subfigure[\label{fig:rate:n2}]{
            \includegraphics[scale=1]{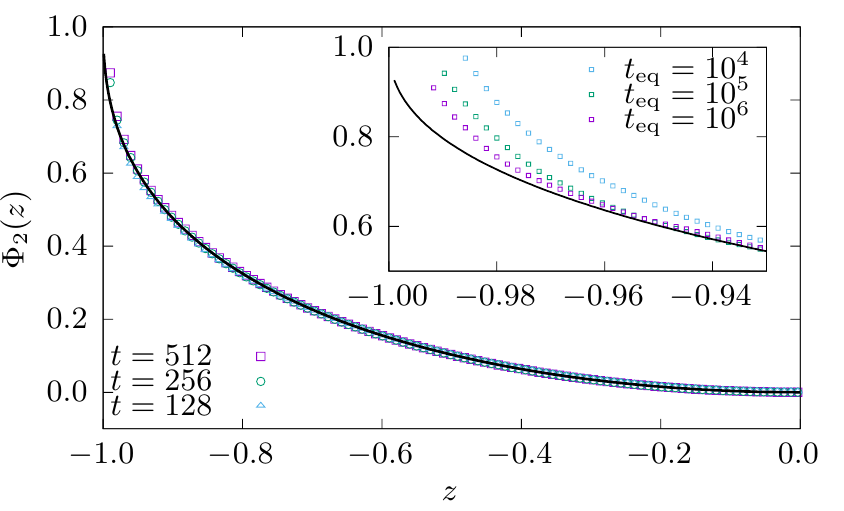}
        }
        \subfigure[\label{fig:rate:n05}]{
            \includegraphics[scale=1]{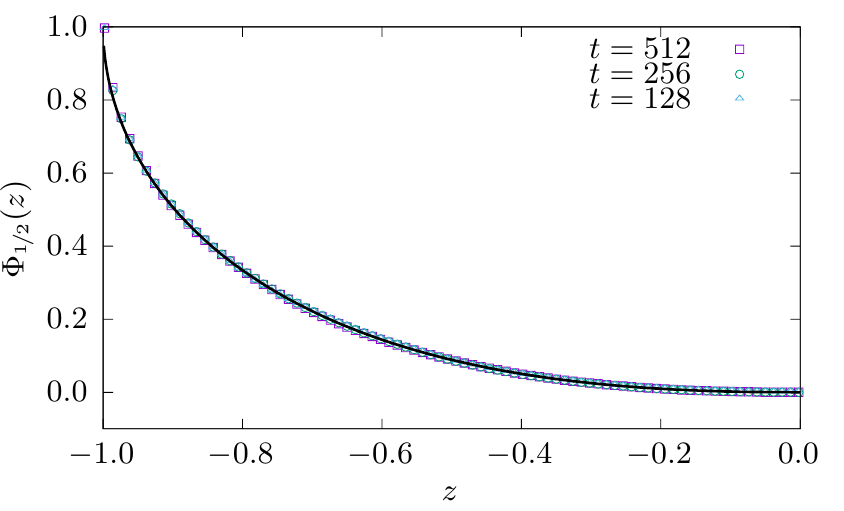}
        }
        \subfigure[\label{fig:rate:n025}]{
            \includegraphics[scale=1]{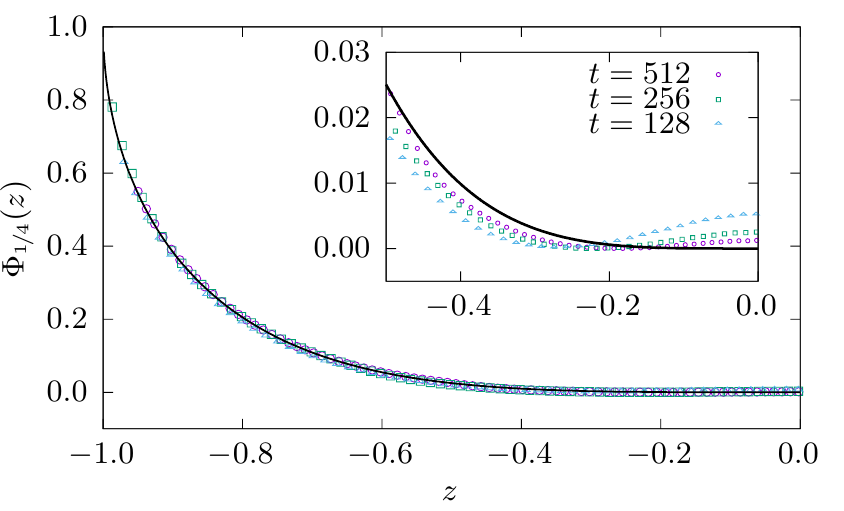}
        }
        \caption{\label{fig:rate}
            Estimates for the rate function from simulation at $t \in \{128, 256, 512\}$
            estimated form Markov chains of the length of $10^7$ sweeps after an
            equilibration time of $10^6$ sweeps at a few dozen different temperatures.
            At these values of $t$, there is almost no size dependence anymore such that we
            assume that the results are already almost converged to the actual rate
            function, which describes the $t\to \infty$ limit.
            The rate function estimates are due to symmetry and to conserve computing
            power only shown for the left branch. All empirical rate functions are shifted
            such that their minimum is at zero.
            (a) The case $n=2$. The inset shows for $t=512$ that higher numerical precision leads
            to results closer to the analytical expectation.
            (b) The $n=1/2$ case. The same slight deviation at the very end of the
            tail occurs as for the $n=2$ case.
            (c) The case $n=1/4$. The inset shows a zoom to illustrate the convergence
            of the two peak structure of $p_n(x, t)$ towards the flat shape of the rate function
            $\Phi_n(z)$.
        }
    \end{figure}

    Obtaining an estimate for the rate function $\Phi_n(z)$, where
    $z=x/x^*(t)$ with $x^*(t)= v_0\, t^n/\Gamma(n+1)$, is more
    complicated, since we need to have high precision data for the tails of
    $p_n(x, t)$. We obtain those for multiple
    values of $t$ using the Markov chain Monte Carlo method described above
    and from this calculate \emph{empirical} rate functions and compare them to the expressions
    from Eqs.~\eqref{phinz_sum_final} and \eqref{phinzb_sum_final} in Fig.~\ref{fig:rate}.
    The symbols represent the results of our Monte Carlo simulations and the lines
    are obtained by numerical maximization of
    Eqs.~\eqref{phinz_sum_final} and \eqref{phinzb_sum_final}. Note that the values of
    $t$ we simulated are already large enough to coincide with the
    apparent asymptotic form within statistical precision, since all values collapse
    onto the asymptotic form.

    We see a very good agreement over most of the support of the rate function.
    The slight deviation from the predicted rate function in the extreme tail
    is caused by difficulties to reach equilibrium, caused by this extremely
    steep tail of extremely rare events.
    In the inset of Fig.~\ref{fig:rate}\subref{fig:rate:n2} we show results obtained with different
    numerical efforts, after different equilibration times $t_\mathrm{eq}$ of
    the Markov chain measured in sweeps, i.e.~$t$ change attempts. From this we estimate
    that we can not reach equilibrium for $z > 0.95$, up to where our two largest
    simulations coincide. However the numerical data for lower values of $z$ should be of
    good quality.
    The very good agreement over most of the rate function confirms the predicted analytical form.

    In Fig.~\ref{fig:rate}\subref{fig:rate:n05} we show the same comparison for the $n=1/2$ case,
    which confirms our analytical rate function convincingly.
    Also here, we observe the very slight deviation at the very end of the tail,
    which arises in the same way as for the $n=2$ case.

    In a similar fashion, our results for the rate function of $n=1/4$ are shown in
    Fig.~\ref{fig:rate}\subref{fig:rate:n025}.
    In this case we could thoroughly sample whole support as well. In contrast to $n\ge 1/2$ we
    see a clear size dependence of our measurements in the inset, where the double humped
    structure of finite $t$ is visible but approaches the monotonous form of the rate function
    with increasing $t$.

\section{Conclusion}
In this paper we studied a class of stochastic processes,
$x(t)= \int_0^t ds\, (t-s)^{n-1}\, \xi(s)$ indexed
by $n>0$. When the driving noise $\xi(t)$ is an uncorrelated Gaussian
white noise (the so called `passive' process), the position distribution $p_n(x,t)$ of $x(t)$ is trivially
Gaussian at all times for all $n>1/2$. The main purpose of this paper was to
consider the case when the driving noise
$\xi(t)$ is an `active' noise, i.e.,  $\xi(t)= (v_0/\Gamma(n))\,\sigma(t)$, where
$\sigma(t)$ is a telegraphic noise
switching between two values $\pm 1$ at a constant rate
$\gamma$. Unlike in the passive case, the position distribution $p_n(x,t)$ in the active case
is well defined for all $n>0$, is highly non-Gaussian and nontrivial to compute.
For $n=1$, our process reduces to the standard run and tumble
process in one dimension. By computing the mean squared displacement exactly,
we found that a localised phase emerges for $0<n<1/2$ where the
variance approaches a constant at late times. This localised phase
owes its origin to the finite memory of the active noise, and has no analogue
in the corresponding passive white noise driven process. For $n> 1/2$,
the variance grows at late times as $\sim t^{n-1/2}$.
In the critical case $n=1/2$, the varies grows slowly as $\ln t$ at late
times.

This localisation transition is also confirmed from the
study of the position distribution $p_n(x,t)$ at late times.
We have shown that for $n<1/2$, the position
distribution approaches a stationary form with a double-humped structure
at late times, while for $n\ge 1/2$ the distribution remains time dependent
even at late times. Furthermore, we have shown that
that the
tails of the position distribution can be described by
the large deviation form: $p_n(x,t)\sim \exp\left[-\gamma\, t\, \Phi_n
\left(\frac{x}{x^*(t)}\right)\right]$ where $x^*(t)= v_0\, t^n/\Gamma(n+1)$.
We computed the large deviation function $\Phi_n(z)$ analytically
for all $n$ and verified it numerically using an importance sampling
algorithm. One of the predictions of our exact computation is
that in the localised phase $n<1/2$, the stationary distribution
has super-exponential tails: $p_n(x,\infty)\sim
\exp\left[-b_n\, |x|^{1/n}\right]$ ($b_n$ being a constant)
as $|x|\to \infty$. Computing the full stationary distribution
$p_n(x,\infty)$ for all $x$ for $n<1/2$, in particular
an analytical description of the double-humped structure,
remains a challenging open problem.

In this paper, we have restricted ourselves only to the one point
function $p_n(x,t)$ of the process $x(t)$. It would be interesting
to compute the multi-time correlation functions,
as well as other observables such as the first-passage probability
for the process $x(t)$ for general $n>0$.

\appendix

\section{Exact cumulant generating function for specific values of $n$}

In this Appendix, we show that for certain specific values of $n$,
namely $n=1$, $n=2$ and the critical case $n=1/2$,
the cumulant generating
function $U^{(n)}(\mu;\, t)$ of the
position distribution $p_n(x,t)$ can be computed exactly.

\subsection{The case $n=1$}

Even though the exact result for $n=1$ is already known in the literature
as mentioned in the introduction, we show
here, for the sake of completeness, how the result follows from the formalism presented in the paper.
Putting $n=1$ in Eq. (\ref{ueq}) gives
\begin{equation}
\frac{d^2 U^{(1)}}{dt^2}+ 2\, \gamma\, \frac{dU^{(1)}}{dt}- \mu^2\, v_0^2\, U^{(1)}=0\, ,
\label{U1eq_.A1}
\end{equation}
to be solved with the boundary conditions: (i) $U^{(1)}(t=0;\,\mu)=1$ and (ii) $dU^{(1)}/dt\Big|_{t=0}=0$.
Since this is a second order differential equation with constant coefficients, the solution can
be trivially obtained in the form $U^{(1)}(t;\, \mu)= A_1\, e^{\lambda\, t}+ A_2\, e^{-\lambda\, t}$
with $\lambda= -\gamma + \sqrt{\gamma^2+ \mu^2\, v_0^2}$.
Upon fixing the two unknown constants via the two boundary conditions we get
\begin{equation}
U^{(1)}(t\, ; \mu)= \frac{1}{2}\left[1+ \frac{\gamma}{\sqrt{\gamma^2+ \mu^2\, v_0^2}}\right]\,
e^{\left(\sqrt{\gamma^2+\mu^2\, v_0^2}-\gamma\right)\, t}
+\frac{1}{2}\left[1- \frac{\gamma}{\sqrt{\gamma^2+ \mu^2\, v_0^2}}\right]\,
e^{-\left(\sqrt{\gamma^2+\mu^2\, v_0^2}+\gamma\right)\, t}\, .
\label{U1sol.A1}
\end{equation}
In the long time limit, the first term dominates over the second term.
Ignoring pre-exponential factors, one finds that as $t\to \infty$
\begin{equation}
U^{(1)}(t\, ; \mu)\sim \exp\left[-\gamma\, t\,
H_1\left(w= \frac{\mu\, v_0}{\gamma}\right)\right]\, ; \quad {\rm with}\quad
H_1(w)= 1- \sqrt{1+w^2}\, ,
\label{h1w.A1}
\end{equation}
in agreement with \eqref{hnw1}. For finite $t$,
by inverting \eqref{U1sol.A1} with respect to $\mu$, one recovers
the known exact distribution $p_1(x,t)$ mentioned
in Eq. \eqref{p1xt.1} of the introduction.

\subsection{The case $n=2$}

In this case, setting $n=2$ in \eqref{ueq} gives the second order differential equation
\begin{equation}
\frac{d^2 U^{(2)}}{dt^2} +\left(2\gamma- \frac{1}{t}\right)\,
\frac{d U^{(2)}}{dt} -\mu^2\, v_0^2\, t^2\, U^{(2)}=0 \, .
\label{u2_eq.A2}
\end{equation}
However, we did not succeed in finding the two linearly independent solutions of this equation in terms of standard
special functions. Instead, we found that for $n=2$, we could solve the separate differential equations
\eqref{U+only.n} and \eqref{U-only.n} for $U_+^{(2)}$ and $U_{-}^{(2)}$ and then add them up to
express the exact solution for $U^{(2)}(t; \mu)$ in terms of standard special functions, in this
case the parabolic cylinder functions.

To proceed, Eqs. (\ref{U+only.n}) and (\ref{U-only.n}) for $n=2$ read
\begin{eqnarray}
\frac{d^2U_+^{(2)}}{dt^2}+ 2\, \gamma\, \frac{dU_+^{(2)}}{dt}+
(\mu\, v_0- \mu^2\, v_0^2\, t^2)\, U_+^{(2)} &= & 0 \label{U+only_2.A2} \\
\frac{d^2U_-^{(2)}}{dt^2}+ 2\, \gamma\, \frac{dU_-^{(2)}}{dt}+
(-\mu\, v_0- \mu^2\, v_0^2\, t^2)\, U_-^{(2)}& =& 0 \, , \label{U-only_2.A2}
\end{eqnarray}
to be solved with the boundary conditions
\begin{equation}
U_{\pm}^{(2)}(t=0;\, \mu)=\frac{1}{2}\, ; \quad {\rm and} \quad \frac{dU_{\pm}^{(2)}}{dt}\Big|_{t=0}=0\, .
\label{ic_1.1}
\end{equation}
It is clear that the solutions satisfy the
following symmetry
\begin{equation}
U_{+}^{(2)}(t;\, \mu)= U_{-}^{(2)}(t;\, -\mu)\, .
\label{symmetry.1}
\end{equation}
Below, we will compute $U_{\pm}(t;\, \mu)$ explicitly by assuming $\mu\ge 0$.
The solution for negative $\mu$ can then be obtained from the symmetry relation \eqref{symmetry.1}

To proceed further, we make the following transformations (assuming $\mu\ge 0$)
\begin{equation}
U_{\pm}^{(2)}(t;\, \mu)= e^{-\gamma\, t}\, \psi_{\pm}\left(\sqrt{2\,\mu\, v_0}\, t\right)\,
\label{U_trans.1}
\end{equation}
that bring the pair of differential equations \eqref{U+only_2.A2} and \eqref{U-only_2.A2}
into a more recognisable form.
We find that $\psi_{\pm}(z)$ satisfy the following differential equations
\begin{eqnarray}
\frac{d^2 \psi_+(z)}{dz^2} +\left(-\frac{\gamma^2}{2\,\mu\, v_0}+
\frac{1}{2}- \frac{z^2}{4}\right)\, \psi_+(z) &=& 0
\label{phi+.1} \\
\frac{d^2 \psi_-(z)}{dz^2} +\left(-\frac{\gamma^2}{2\,\mu\, v_0}-
\frac{1}{2}- \frac{z^2}{4}\right)\, \psi_-(z) &=& 0
\, .
\label{phi-.1}
\end{eqnarray}
These equations resemble the Schr\"odinger equations for a harmonic oscillator potential. Indeed, the differential equation
\begin{equation}
\frac{d^2 f}{dz^2} + \left(p+\frac{1}{2}- \frac{z^2}{4}\right)\, f(z)=0
\label{diff_eq.1}
\end{equation}
has two linearly independent solutions $D_p(z)$ and $D_p(-z)$ known as parabolic cylinder functions~\cite{GR}.
Hence, identifying $p= -\gamma^2/(2\, \mu\, v_0)=-q$ we can write the most
general solutions for $U_{\pm}^{(2)}(t;\, \mu)$
as follows.

For $U_+^{(2)}(t; \, \mu)$ we get
\begin{equation}
U_+^{(2)}(t;\mu)= e^{-\gamma\, t}\left[ A_1(q)\, D_{-q}\left(\sqrt{2\, \mu\, v_0}\,\, t\right)
+A_2(q)\, D_{-q}\left(-\sqrt{2\, \mu\, v_0}\,\, t\right)\right]\, ;  \quad {\rm with}\,\,
q= \frac{\gamma^2}{2\,\mu\, v_0}\ge 0\, ,
\label{Uplus_sol.1}
\end{equation}
where $A_1(q)$ and $A_2(q)$ are two arbitrary constants to be fixed from the initial conditions
in \eqref{ic_1.1}. The
two initial conditions $U_+^{(2)}(0; \, \mu)=1/2$ and $ {\dot U}_+^{(2)}(t=0;\, \mu)=0$
give two relations between
$A_1(q)$ and $A_2(q)$
\begin{eqnarray}
A_1(q) + A_2(q) & = &\frac{1}{2\, D_{-q}(0)} \, \\
A_1(q) -A_2(q)   &=&  \frac{\gamma}{2\, \sqrt{2\, \mu\, v_0}\, D_{-q}'(0)}\, ,
\label{lin_rel.1}
\end{eqnarray}
where $D_{-q}'(z)= d D_{-q}(z)/dz\Big|_{z=0}$. These two relations fix the two constants
\begin{eqnarray}
A_1(q) & =& \frac{1}{4}\left( \frac{1}{D_{-q}(0)}+
\frac{\gamma}{ \sqrt{2\, \mu\, v_0}\, D_{-q}'(0)}\right)=
\frac{2^{q/2}}{4\,\sqrt{\pi}}\, \left[\Gamma((1+q)/2)-\sqrt{\frac{q}{2}}\, \Gamma(q/2)\right]
\label{A1.1}\\
A_2(q) &= & \frac{1}{4}\left( \frac{1}{D_{-q}(0)}-
\frac{\gamma}{ \sqrt{2\, \mu\, v_0}\, D_{-q}'(0)}\right)=
\frac{2^{q/2}}{4\,\sqrt{\pi}}\, \left[\Gamma((1+q)/2)+\sqrt{\frac{q}{2}}\, \Gamma(q/2)\right]
\, .
\label{A2.1}
\end{eqnarray}
where we used the explicit values of $D_{-q}(0)$ and $D_{-q}'(0)$~\cite{GR}.

Similarly, we can find the solution for $U_{-}^{(2)}(t;\, \mu)$ (again for $\mu\ge 0$)
\begin{equation}
U_-^{(2)}(t;\mu)= e^{-\gamma\, t}\left[ B_1(q)\, D_{-q-1}\left(\sqrt{2\, \mu\, v_0}\,\, t\right)
+B_2(q)\, D_{-q-1}\left(-\sqrt{2\, \mu\, v_0}\,\, t\right)\right]\, ;  \quad {\rm with}\,\,
q= \frac{\gamma^2}{2\,\mu\, v_0}\ge 0\, ,
\label{Uminus_sol.1}
\end{equation}
where the two constants $B_1(q)$ and $B_2(q)$ are given by
\begin{eqnarray}
B_1(q) & =& \frac{1}{4}\left( \frac{1}{D_{-q-1}(0)}+
\frac{\gamma}{ \sqrt{2\, \mu\, v_0}\, D_{-q-1}'(0)}\right)
= -\sqrt{q}\, \frac{2^{q/2}}{4\,\sqrt{\pi}}\,
\left[\Gamma((1+q)/2)-\sqrt{\frac{q}{2}}\, \Gamma(q/2)\right]= -\sqrt{q}\, A_1(q) \label{B1.1} \\
B_2(q) &= & \frac{1}{4}\left( \frac{1}{D_{-q-1}(0)}-
\frac{\gamma}{ \sqrt{2\, \mu\, v_0}\, D_{-q-1}'(0)}\right)
= \sqrt{q}\, \frac{2^{q/2}}{4\,\sqrt{\pi}}\, \left[\Gamma((1+q)/2)+\sqrt{\frac{q}{2}}\, \Gamma(q/2)\right]
= \sqrt{q}\, A_2(q)\, .
\label{B2.1}
\end{eqnarray}
Indeed, by replacing $q\to -q$, one can now verify explicitly that the symmetry relation
in \eqref{symmetry.1} is satisfied by
these solutions $U_{\pm}^{(2)}(t;\, \mu)$.

Finally, the total cumulant generating function for any $\mu$ is given by the sum
\begin{eqnarray}
U^{(2)}(t;\, \mu)&= & U_{+}^{(2)}(t;\, \mu)+U_{-}^{(2)}(t;\, \mu)=
\int_{-\infty}^{\infty} e^{-\mu\, x}\, p_2(x,t)\, dx
\nonumber \\
&=& e^{-\gamma\, t}\left[A_1(q)\,D_{-q}\left(\frac{\gamma}{\sqrt{q}}\, t\right)
+A_2(q)\, D_{-q}\left(-\frac{\gamma}{\sqrt{q}}\,\, t\right)-\sqrt{q}\, A_1(q)\,
D_{-q-1}\left(\frac{\gamma}{\sqrt{q}}\,\, t\right)
+\sqrt{q}\, A_2(q)\, D_{-q-1}\left(-\frac{\gamma}{\sqrt{q}}\, t\right)\right] \nonumber \\
\label{Ut_final.1}
\end{eqnarray}
where $ q= \gamma^2/(2\,\mu\, v_0)$ and the two constants $A_1(q)$ and $A_2(q)$
are given respectively in \eqref{A1.1} and \eqref{A2.1}. Manifestly, we have the symmetry
$U^{(2)}(t;\, \mu)= U^{(2)}(t;\, -\mu)$.

We now show briefly how to recover the large deviation behaviour at late times in
\eqref{Ut_ldp.n2} for $n=2$ from the exact solution in \eqref{Ut_final.1}.
To proceed, we first rewrite the exact solution \eqref{Ut_final.1}
in terms
of $w= \mu\, v_0\, t/(2\,\gamma)$, i.e., replace $t= 2\,\gamma\, w/(\mu\, v_0)$ and
use $q= \gamma^2/(2\,\mu\, v_0)$. This gives
\begin{equation}
U(t;\,\mu)=  e^{-\gamma\, t}\left[A_1(q)\,D_{-q}\left(4\,\sqrt{q}\, w\right)
+A_2(q)\, D_{-q}\left(-4\, \sqrt{q}\, w\right)-\sqrt{q}\, A_1(q)\,
D_{-q-1}\left(4\, \sqrt{q}\, w\right)
+\sqrt{q}\, A_2(q)\, D_{-q-1}\left(-4\, \sqrt{q}\, w\right)\right]\, .
\label{Ut_final.2}
\end{equation}
We will henceforth assume that $\mu> 0$, i.e., $q>0$ and $w>0$. For $\mu< 0$, the
result will follow from the symmetry $U^{(2)}(t;\, \mu)= U^{(2)}(t;\, -\mu)$,
which also indicates the symmetry $H_2(w)=H_2(-w)$.

Since $q= \gamma^2/(2\,\mu\, v_0)\to \infty$ as $\mu\to 0$, we need to analyse
\eqref{Ut_final.2} in the limit $q\to \infty$ keeping $w$ fixed.
To take this tricky limit where both the index and the argument of the parabolic
cylinder function diverge, we found it convenient to use the following
integral representation
of $D_{-q}(z)$~\cite{GR}, valid for $q>0$
\begin{equation}
D_{-q}(z)=\frac{e^{-z^2/4}}{\Gamma(q)}\, \int_0^{\infty}
e^{-x\, z- x^2/2+ (q-1)\, \ln x}\, dx \, .
\label{int_rep.1}
\end{equation}
Using this representation, we now evaluate each of the four terms on the
right hand side (r.h.s)
of \eqref{Ut_final.2} in the $q\to \infty$ limit.
For instance, the first term
(excluding the global $e^{-\gamma\,t}$ factor) can be written as
\begin{equation}
A_1(q)\, \,D_{-q}\left(4\,\sqrt{q}\, w\right)=
\frac{A_1(q)}{\Gamma(q)}\, e^{-4\, q\, w^2}\,
\int_{0}^{\infty} e^{- 4\,\sqrt{q}\, x\, w- x^2/2+ (q-1)\, \ln x}\, dx\, ,
\label{t1.1}
\end{equation}
where $A_1(q)$ in given in \eqref{A1.1}.
In the large $q$ limit, the integral can be evaluated by the saddle point method. Using
also the asymptotic expansion of $\Gamma(q)$ for large $q$, we find, up to pre-exponential
factors
\begin{equation}
A_1(q)\, \,D_{-q}\left(4\,\sqrt{q}\, w\right)\sim e^{q\, f_1(w)}\, ; \quad {\rm where}\quad
f_1(w)= - 2 \, w\, \sqrt{1+4\, w^2} + \ln \left(\sqrt{1+4\, w^2}-2\, w\right)\, .
\label{f1w.1}
\end{equation}
Similarly, the second term gives in the large $q$ limit
\begin{equation}
A_2(q)\, \,D_{-q}\left(-4\,\sqrt{q}\, w\right)\sim e^{q\, f_2(w)}\, ; \quad {\rm where}\quad
f_2(w)=  2 \, w\, \sqrt{1+4\, w^2} + \ln \left(\sqrt{1+4\, w^2}+2\, w\right)\, .
\label{f2w.1}
\end{equation}
Likewise, one can show that the third term scales as $\sim e^{q\, f_1(w)}$, while the
fourth term scales as $\sim e^{q\, f_2(w)}$. Now, it is easy to see that for all $w>0$
(we recall that $w= \mu\, v_0\, t/(2\, \gamma)>0$ for $\mu>0$), $f_2(w)>f_1(w)$. Hence,
for large $q$ and $w>0$, the r.h.s of \eqref{Ut_final.2} scales as
\begin{equation}
U(t;\,\mu)\sim  e^{-\gamma\, t}\, e^{q\, f_2(w)} \sim
e^{-\gamma\, t\, \left(1- \frac{f_2(w)}{4w}\right)}
\label{Ut_scaling.1}
\end{equation}
where we replaced $q=\gamma\, t/(4\, w)$.
From \eqref{Ut_scaling.1} we read off $H_2(w)$ for $w>0$
\begin{equation}
H_2(w)= 1-\frac{f_2(w)}{w}= 1- \frac{1}{2}\, \sqrt{1+ 4\, w^2}- \frac{1}{4\, w}\, \ln \left(
\sqrt{1+4\, w^2}+2\, w\right)\, ; \quad {\rm for}\,\, w\ge 0\, .
\label{Hw.1}
\end{equation}
Note that the result for $H_2(w)$ with $w<0$ just follows from the symmetry
$H_2(-w)=H_2(w)$. In fact, the function $H_2(w)$ in \eqref{Hw.1}
can be written in a manifestly symmetric form that is valid for all $-\infty<w<\infty$ as
\begin{equation}
H_2(w)= 1- \frac{1}{2}\, \sqrt{1+ 4\, w^2}-
\frac{{\rm arcsinh}(2\, w)}{4\, w}\, ;
\quad -\infty<w<\infty \, .
\label{Hw_symmetric}
\end{equation}
This result coincides exactly with \eqref{hnw2} derived in Section \ref{ld}.

\subsection{The case $n=1/2$}

In this case, we start from \eqref{ueq} which for $n=1/2$ reads
\begin{equation}
\frac{d^2 U^{(1/2)}}{dt^2} +\left(2\gamma+ \frac{1}{2\,t}\right)\,
\frac{d U^{(1/2)}}{dt} -\frac{\mu^2\, v_0^2}{\pi\,t}\,
U^{(1/2)}=0 \,
\label{ueq_nhalf}
\end{equation}
to be solved with the initial conditions in \eqref{unhalf_bc.1}.
To reduce it to a familiar differential equation we make the following
transformation
\begin{equation}
U^{(1/2)}(t;\mu)= \sqrt{2\,\gamma\,t}\, e^{-2\, \gamma\, t}\, F(2\,\gamma\,t)\, .
\label{U2F_trans.1}
\end{equation}
Substituting \eqref{U2F_trans.1} in \eqref{ueq_nhalf}, one can check
that $F(z)$ satisfies the differential equation
\begin{equation}
z\, F''(z)+ \left(\frac{3}{2}-z\right)\, F'(z)- \left(1+
\frac{\mu^2\, v_0^2}{2\,\pi\, \gamma}\right)\, F(z)=0\, .
\label{Fzeq.1}
\end{equation}
This is now of the standard form of the
confluent hypergeometric differential equation~\cite{abr65}:
$F''(z)+ (b-z)F'(z)-a F(z)=0$, whose
general solution
is given by a linear combination of two independent
confluent hypergeometric functions $U(a,b,z)$ and ${}_1F_1(a,b,z)$.
Hence,
\begin{equation}
F(z)= C_1\, U\left(1+ \frac{\mu^2\, v_0^2}{2\,\pi\, \gamma}, \frac{3}{2},z\right) +
C_2\,\, {}_1F_1\left(1+ \frac{\mu^2\, v_0^2}{2\,\pi\, \gamma},\frac{3}{2},z
\right)\, ,
\label{gen_sol_FZ.1}
\end{equation}
where $C_1$ and $C_2$ are unknown constants.
Hence, the general solution of \eqref{ueq_nhalf} can be written as
\begin{equation}
U^{(1/2)}(t;\mu)= \sqrt{2\,\gamma\,t}\, e^{-2\,\gamma\, t}\,
\left[ C_1\, U\left(1+ \frac{\mu^2\, v_0^2}{2\,\pi\, \gamma}, \frac{3}{2},
2\,\gamma\, t\right) + C_2\,\,
{}_1F_1\left(1+ \frac{\mu^2\, v_0^2}{2\,\pi\, \gamma},\frac{3}{2},
2\,\gamma\, t\right)\right]\, .
\label{gen_sol_Ut.1}
\end{equation}
The unknown constants are fixed from the initial conditions
in \eqref{unhalf_bc.1}. This gives
\begin{equation}
C_1= \frac{\Gamma\left(1+ \frac{\mu^2\, v_0^2}{2\,\pi\, \gamma}
\right)}{\sqrt{\pi}}\, ;
\quad {\rm and}\quad
C_2= 2\, \frac{\Gamma\left(1+ \frac{\mu^2\, v_0^2}{2\,\pi\,
\gamma}\right)}{\Gamma\left(\frac{1}{2}+ \frac{\mu^2\, v_0^2}{2\,\pi\,
\gamma}\right)}\, .
\label{c1c2.1}
\end{equation}
Hence the final exact cumulant generating function for $n=1/2$ is given by
\begin{equation}
U^{(1/2)}(t;\mu)= \sqrt{2\,\gamma\,t}\, e^{-2\,\gamma\, t}\,
\Gamma\left(1+ \frac{\mu^2\, v_0^2}{2\,\pi\,\gamma}\right)\,
\left[ \frac{1}{\sqrt{\pi}}\, U\left(1+ \frac{\mu^2\, v_0^2}{2\,\pi\, \gamma}, \frac{3}{2},
2\,\gamma\, t\right)
+ \frac{2}{\Gamma\left(\frac{1}{2}+
\frac{\mu^2\, v_0^2}{2\,\pi\,\gamma}\right)}\,
{}_1F_1\left(1+ \frac{\mu^2\, v_0^2}{2\,\pi\, \gamma},\frac{3}{2},
2\,\gamma\, t\right)\right]\, .
\label{nhalf_final.1}
\end{equation}
In the limit $t\to \infty$, $\mu\to \infty$ but keeping
$w= 2\,\mu\,v_0/{\gamma\, \sqrt{\pi t}}$ fixed, we expect
that this exact solution should converge to the large deviation form
\begin{equation}
U^{(1/2)}(t;\mu) \sim
\exp\left[- \gamma\, t\, H_{1/2}\left(w=\frac{2\,\mu\, v_0}{\gamma\,
\sqrt{\pi t}}\right)\right]\, ,
\label{U12_expect.1}
\end{equation}
where the rate function $H_{1/2}(w)$ is given in \eqref{h_halfw.1}.
We have not proved it here, but we have checked by Mathematica that
indeed the exact solution \eqref{nhalf_final.1}
does converge to this expected large deviation form in \eqref{U12_expect.1}.

\end{document}